\documentstyle[12pt,psfig]{article}

\newcommand{\ra}{\rightarrow}

\newcommand{\be}{\begin{equation}}
\newcommand{\ee}{\end{equation}}
\newcommand{\bea}{\begin{eqnarray}}
\newcommand{\eea}{\end{eqnarray}}
\newtheorem{lem}{Lemma}[section]
\newtheorem{pro}{Proposition}[section]
\newtheorem{thm}{Theorem}[section]

\renewcommand{\thefootnote}{\alph{footnote}}

\def\bbbr{{\rm I\!R}} 

\def\bbbn{{\rm I\!N}} 

\def\bbbm{{\rm I\!M}}

\def\bbbh{{\rm I\!H}}

\def\bbbf{{\rm I\!F}}

\def\bbbk{{\rm I\!K}}

\def\bbbl{{\rm I\!L}}

\def\bbbp{{\rm I\!P}}

\def\bbbe{{\rm I\!E}}

\def\bbbone{{\mathchoice {\rm 1\mskip-4mu l} {\rm 1\mskip-4mu l}
{\rm 1\mskip-4.5mu l} {\rm 1\mskip-5mu l}}}

\def\bbbc{{\mathchoice {\setbox0=\hbox{$\displaystyle\rm C$}\hbox{\hbox
to0pt{\kern0.4\wd0\vrule height0.9\ht0\hss}\box0}}
{\setbox0=\hbox{$\textstyle\rm C$}\hbox{\hbox
to0pt{\kern0.4\wd0\vrule height0.9\ht0\hss}\box0}}
{\setbox0=\hbox{$\scriptstyle\rm C$}\hbox{\hbox
to0pt{\kern0.4\wd0\vrule height0.9\ht0\hss}\box0}}
{\setbox0=\hbox{$\scriptscriptstyle\rm C$}\hbox{\hbox
to0pt{\kern0.4\wd0\vrule height0.9\ht0\hss}\box0}}}}

\def\bbbq{{\mathchoice {\setbox0=\hbox{$\displaystyle\rm Q$}\hbox{\raise
0.15\ht0\hbox to0pt{\kern0.4\wd0\vrule height0.8\ht0\hss}\box0}}
{\setbox0=\hbox{$\textstyle\rm Q$}\hbox{\raise
0.15\ht0\hbox to0pt{\kern0.4\wd0\vrule height0.8\ht0\hss}\box0}}
{\setbox0=\hbox{$\scriptstyle\rm Q$}\hbox{\raise
0.15\ht0\hbox to0pt{\kern0.4\wd0\vrule height0.7\ht0\hss}\box0}}
{\setbox0=\hbox{$\scriptscriptstyle\rm Q$}\hbox{\raise
0.15\ht0\hbox to0pt{\kern0.4\wd0\vrule height0.7\ht0\hss}\box0}}}}
\def\bbbt{{\mathchoice {\setbox0=\hbox{$\displaystyle\rm
T$}\hbox{\hbox to0pt{\kern0.3\wd0\vrule height0.9\ht0\hss}\box0}}
{\setbox0=\hbox{$\textstyle\rm T$}\hbox{\hbox
to0pt{\kern0.3\wd0\vrule height0.9\ht0\hss}\box0}}
{\setbox0=\hbox{$\scriptstyle\rm T$}\hbox{\hbox
to0pt{\kern0.3\wd0\vrule height0.9\ht0\hss}\box0}}
{\setbox0=\hbox{$\scriptscriptstyle\rm T$}\hbox{\hbox
to0pt{\kern0.3\wd0\vrule height0.9\ht0\hss}\box0}}}}

\def\bbbs{{\mathchoice
{\setbox0=\hbox{$\displaystyle     \rm S$}\hbox{\raise0.5\ht0\hbox
to0pt{\kern0.35\wd0\vrule height0.45\ht0\hss}\hbox
to0pt{\kern0.55\wd0\vrule height0.5\ht0\hss}\box0}}
{\setbox0=\hbox{$\textstyle        \rm S$}\hbox{\raise0.5\ht0\hbox
to0pt{\kern0.35\wd0\vrule height0.45\ht0\hss}\hbox
to0pt{\kern0.55\wd0\vrule height0.5\ht0\hss}\box0}}
{\setbox0=\hbox{$\scriptstyle      \rm S$}\hbox{\raise0.5\ht0\hbox
to0pt{\kern0.35\wd0\vrule height0.45\ht0\hss}\raise0.05\ht0\hbox
to0pt{\kern0.5\wd0\vrule height0.45\ht0\hss}\box0}}
{\setbox0=\hbox{$\scriptscriptstyle\rm S$}\hbox{\raise0.5\ht0\hbox
to0pt{\kern0.4\wd0\vrule height0.45\ht0\hss}\raise0.05\ht0\hbox
to0pt{\kern0.55\wd0\vrule height0.45\ht0\hss}\box0}}}}

\def\bbbz{{\mathchoice {\hbox{$\sf\textstyle Z\kern-0.4em Z$}}
{\hbox{$\sf\textstyle Z\kern-0.4em Z$}}
{\hbox{$\sf\scriptstyle Z\kern-0.3em Z$}}
{\hbox{$\sf\scriptscriptstyle Z\kern-0.2em Z$}}}}

\def\bC{\ifmmode{\bbbc }\else${\bbbc }$\fi}
\def\bE{\ifmmode{\bbbe }\else${\bbbe }$\fi}
\def\bF{\ifmmode{\bbbf }\else${\bbbf }$\fi}
\def\bH{\ifmmode{\bbbh }\else${\bbbh }$\fi}
\def\bI{\ifmmode{\bbbone }\else${\bbbone }$\fi}
\def\bK{\ifmmode{\bbbk }\else${\bbbk }$\fi}
\def\bL{\ifmmode{\bbbl }\else${\bbbl }$\fi}
\def\bM{\ifmmode{\bbbm }\else${\bbbm }$\fi}
\def\bN{\ifmmode{\bbbn }\else${\bbbn }$\fi}
\def\bP{\ifmmode{\bbbp }\else${\bbbp }$\fi}
\def\bQ{\ifmmode{\bbbq }\else${\bbbq }$\fi}
\def\bR{\ifmmode{\bbbr }\else${\bbbr }$\fi}
\def\bS{\ifmmode{\bbbs }\else${\bbbs }$\fi}
\def\bZ{\ifmmode{\bbbz }\else${\bbbz }$\fi}

\def\cA{\ifmmode{\cal A}\else${\cal A}$\fi}
\def\cB{\ifmmode{\cal B}\else${\cal B}$\fi}
\def\cC{\ifmmode{\cal C}\else${\cal C}$\fi}
\def\cD{\ifmmode{\cal D}\else${\cal D}$\fi}
\def\cE{\ifmmode{\cal E}\else${\cal E}$\fi}
\def\cF{\ifmmode{\cal F}\else${\cal F}$\fi}
\def\cG{\ifmmode{\cal G}\else${\cal G}$\fi}
\def\cH{\ifmmode{\cal H}\else${\cal H}$\fi}
\def\cI{\ifmmode{\cal I}\else${\cal I}$\fi}
\def\cJ{\ifmmode{\cal J}\else${\cal J}$\fi}
\def\cK{\ifmmode{\cal K}\else${\cal K}$\fi}
\def\cL{\ifmmode{\cal L}\else${\cal L}$\fi}
\def\cM{\ifmmode{\cal M}\else${\cal M}$\fi}
\def\cN{\ifmmode{\cal N}\else${\cal N}$\fi}
\def\cO{\ifmmode{\cal O}\else${\cal O}$\fi}
\def\cP{\ifmmode{\cal P}\else${\cal P}$\fi}
\def\cQ{\ifmmode{\cal Q}\else${\cal Q}$\fi}
\def\cR{\ifmmode{\cal R}\else${\cal R}$\fi}
\def\cS{\ifmmode{\cal S}\else${\cal S}$\fi}
\def\cT{\ifmmode{\cal T}\else${\cal T}$\fi}
\def\cU{\ifmmode{\cal U}\else${\cal U}$\fi}
\def\cV{\ifmmode{\cal V}\else${\cal V}$\fi}
\def\cW{\ifmmode{\cal W}\else${\cal W}$\fi}
\def\cX{\ifmmode{\cal X}\else${\cal X}$\fi}
\def\cY{\ifmmode{\cal Y}\else${\cal Y}$\fi}
\def\cZ{\ifmmode{\cal Z}\else${\cal Z}$\fi}

\def\bk#1{{\ifmmode{\langle#1\rangle}\else${\langle#1\rangle}$\fi}} 

\def\sp#1#2{{\ifmmode{\langle{#1}|{#2}\rangle}
\else${\langle{#1}|{#2}\rangle}$\fi}}

\def\prf{\medbreak\noindent{\bf Proof.}\enspace}

\def\qed{\hfill $\sqcap$\llap{$\sqcup$}\break}

\def\data{\the\day\space\ifcase\month\or January \or February \or March \or
April \or May \or June \or July \or August \or September
\or October \or November \or December \fi\space\the\year}

\def\date{\the\day\space\ifcase\month\or Janvier \or F\'evrier \or Mars \or
Avril \or Mai \or Juin \or Juillet \or Ao\^ut \or Septembre
\or Octobre \or Novembre \or D\'ecembre \fi\space\the\year}

\def\sc{\scriptstyle}

\def\writefig#1 #2 #3 {\rlap{\kern #1 truecm
\raise #2 truecm \hbox{#3}}}
\def\figtext#1{\smash{\hbox{#1}}
\vspace{-5mm}}
\def\math#1{\ifmmode
\mathchoice{\mbox{$\displaystyle\rm#1$}}
{\mbox{$\textstyle\rm#1$}}
{\mbox{$\scriptstyle\rm#1$}}
{\mbox{$\scriptscriptstyle\rm#1$}}\else
{\mbox{$\rm#1$}}\fi}

\def\W{{\tt W}}

\def\stau{\hat{\tau}}
\def\staubd{\hat{\tau}_{\math{\footnotesize bd}}}
\def\req#1{(\ref{#1})}
\def\nnm{\nonumber}

\def\bk#1{\langle#1\rangle}

\def\writefig#1 #2 #3 {\rlap{\kern #1 truecm
\raise #2 truecm \hbox{#3}}}
\def\figtext#1{\smash{\hbox{#1}}
\vspace{-5mm}}

\begin{document}

\null
\vspace{2cm}\noindent
\centerline{{\Large\bf INTERFACE PINNING}}
\newline
\newline
\centerline{{\Large\bf AND FINITE SIZE EFFECTS}}
\newline
\newline
\centerline{{\Large\bf IN THE 2D ISING MODEL}}
\newline
\newline
\centerline{C.-E. Pfister\footnotemark[1],
Y. Velenik\footnotemark[2]}
\newline

\setcounter{footnote}{0}
\centerline{\footnotemark D\'epartement de Math\'ematiques, EPF-L }
\centerline{CH-1015 Lausanne, Switzerland}
\centerline{e-mail: charles.pfister@dma.epfl.ch}
\medskip
\centerline{\footnotemark D\'epartement de Physique, EPF-L}
\centerline{CH-1015 Lausanne, Switzerland}
\centerline{e-mail: velenik@dpmail.epfl.ch}
\renewcommand{\thefootnote}{\arabic{footnote}}
\setcounter{footnote}{0}
\vspace*{2cm}
\centerline{\data}
\noindent

\vspace*{2cm}
\noindent
{\Large\bf Abstract:} 
We apply new techniques  developed in 
\cite{PV1} to the study of some surface effects in the 2D Ising 
model. We examine in particular the pinning-depinning transition. 
The results are valid for all subcritical temperatures. 
By duality we
obtained new finite size effects on
the asymptotic behaviour of the  two--point correlation function 
above the critical temperature. 
The key--point of the analysis is to obtain good concentration properties 
of the measure defined on  the random lines giving the high--temperature 
representation of the two--point correlation function, as a consequence
of the {\bf sharp  triangle inequality}: let $\stau(x)$ be the surface 
tension of an interface perpendicular to $x$; then for any $x$, $y$
$$
\stau(x)+\stau(y)-\stau(x+y)\geq {1\over \kappa}
\left(\|x\|+\|y\|-\|x+y\|\right)\,,
$$
where $\kappa$ is the maximum curvature of the Wulff shape and
$\|x\|$ the Euclidean norm of $x$.

\bigskip
\bigskip


\pagebreak

\markboth{Interface Pinning in the 2D Ising Model}
         {Interface Pinning in the 2D Ising Model}


\section{Introduction}\label{Introduction}
\setcounter{equation}{0}

Consider a 2D Ising model in some rectangular box with boundary
conditions implying the presence of a phase separation line crossing 
the  box from one vertical side to the other one. The bottom side of 
the box, which we call the  wall, is subject to a magnetic field
$h$. By varying $\beta$ and/or $h$, when we are in the phase
coexistence region, we can observe the so-called pinning--depinning
transition,  which occurs when the $+$ phase, which is above the
interface, begins to wet the wall with the result that the
equilibrium shape of the interface changes from a straight line 
crossing the box to a broken line touching a macroscopic part of the
wall. This phenomenon  has been recently studied by
Patrick \cite{Pa} in the  SOS model using exact
calculations. In the 2D Ising model this phenomenon has a dual 
interpretation at high temperature in terms of 
finite size effects on the 
asymptotic behaviour of two--point function for large distances. 
These questions can be analyzed in the 2D Ising model by the new
non--perturbative  results  developed  in \cite{Pf2} and \cite{PV1}
in the context of  large deviations  and  separation of phases. 
Some parts of the paper , like section \ref{Ellipse}, are written
directly for the two--point function. Pinning--depinning transition
and the finite size effects on the two--point correlation function 
are treated in section \ref{Continuum}.

The fundamental thermodynamical function associated with 
an interface is the surface tension. 
The interface\footnote{ The concept of 
``interface'' as a macroscopic phenomenon is advocated in the recent
paper  \cite{ABCP}. Moreover, Talagrand in his analysis \cite{T} about 
the Law of Large Numbers for independent random variables 
develops similar ideas. See also footnote 2.} 
between the two phases of the model is a non--random object.
On the other hand,  at the scale of the lattice spacing,
we have the random line, which is   a
geometrical object separating the two phases.
The interface is
therefore defined  at a scale where the fluctuations of the phase
separation line become negligible. Its main properties  are described
by a functional of the surface tension. The observed interface at
equilibrium is a minimum of this functional
(section \ref{Variational}). The study of
fluctuations of the phase separation line is an
important and difficult problem; some 
works in that directions are \cite{Hi}, 
\cite{BLP1}\footnote{The local structure of the phase separation line
is studied in \cite{BLP1} at low temperatures for the case of the
so--called
$\pm$ boundary condition, which corresponds to $a=b=1/2$ and $h=1$
of the present paper. The definition of the phase
separation line in \cite{BLP1}  coincides with the
one of Gallavotti in his work \cite{G} about the phase
separation in the 2D Ising model; it 
differs slightly from the one used here, but not in an
essential way. (Notice that  the terminology ``interface'' is
sometimes used for ``phase separation line'' in \cite{BLP1}.) It is
shown that the phase separation line has a well-defined intrinsic
width,  which is finite at the scale of the lattice spacing, but
that its position has fluctuations typically of the order of 
$O(L^{1/2})$, $L$ being the linear size of the box $\Lambda_L$
containing the system. Because
of these fluctuations the projection of the corresponding limiting
Gibbs state, at the middle of the box, when $L\ra\infty$, is
translation invariant; the magnetization (at the middle of the box)
is zero. However, the results of this paper show that, at a suitable
mesoscopic scale of order $O(L^\alpha)$, $\alpha>1/2$, the system
has a well--defined non--random horizontal interface. To describe the
system at the scale  $O(L^\alpha)$ we partition the box $\Lambda_L$
into square boxes $C_i$ of linear size $O(L^\alpha)$; the state
of the system in each of these boxes is specified by the empirical
magnetization  $|C_i|^{-1}\sum_{t\in C_i}\sigma(t)$ (normalized
block-spin). Then we rescale all lengths by $1/L$ in order to get a
measure  for these normalized block-spins in the fixed (macroscopic
box) $Q$. When $L\ra\infty$ these measures converge to a non--random
macroscopic configuration with a well-defined horizontal interface
separating the two phases of the model, characterized by a value 
$\pm m^*$ of the normalized block-spins, $m^*$ being the spontaneous
magnetization of the  model.}, \cite{DH}. 

A key--point of the present analysis
is the role of the sharp triangle inequality of the surface 
tension \cite{I}, which, combined with our recent results, leads to
good concentration properties for the measure defined on the random 
lines giving the high--temperature representation of the two--point 
correlation function (section \ref{Ellipse}).
(These random lines coincide with the phase--separation lines.)
Because of its importance, we devote  section \ref{sti}
to a geometrical study of the sharp triangle inequality.
This section can be read independently; Proposition
\ref{stiequivalence} has its own interest.

The paper is not self--contained, because we use in an essential
way  results of \cite{PV1}, in particular those of section 5. 
They are carefully stated in Propositions \ref{prorandom} and
\ref{propv} and Lemmas \ref{lem5.1} and \ref{lem5.2}.
This has the advantage that we can focus our attention on
the  essential points of the proofs.
Motivated by \cite{Pa} we have chosen  pinning--depinning
transition to illustrate the technique of \cite{PV1};
we can consider more 
complicated situations than the ones of this paper.

{\bf Acknowledgments:} We thank M. Troyanov for very useful 
discussions and suggestions about the geometrical aspect of 
section \ref{sti}.


\section{Definitions and notations}\label{Definitions}
\setcounter{equation}{0}

\subsection{Phase separation line}

We follow \cite{PV1} for the notation and terminology. 
Throughout the paper $O(x)$ denotes a
non--negative function of 
$x\in\bR^+$, such that there exists a constant $C$ with $O(x)\leq
Cx$. The function $O(x)$ may be different at different
places.

Let $Q$ be the square box in $\bR^2$,
\be
Q:=\{\,x=(x(1),x(2))\in\bR^2:\,|x(1)|\leq1/2\;,\;0\leq x(2)\leq 1\,\}
\,,
\ee
and $\Lambda_L$ be the subset of $\bZ^2$ ($L$ an even integer)
\be\label{lambda}
\Lambda_L:=\{\,x=(x(1),x(2))\in\bZ^2:\,|x(1)|\leq L/2\;,\;0
\leq x(2)\leq L\,\}\,.
\ee
The spin variable at $x\in\bZ^2$ is the random variable 
$\sigma(x)=\pm 1$; spin configurations are denoted by
$\omega\in\{-1,+1\}^{\bZ^2}$, so that $\sigma(t)(\omega)=\pm 1$
if $\omega(t)=\pm 1$.
 We always  suppose that we have for the
box $\Lambda_L$ either the  $ab$ boundary condition ($ab$ b.c.) or the
$-$ boundary condition ($-$ b.c.).  Let $0\leq a\leq 1$ and $0 \leq b \leq 1$ be given;
the $ab$ b.c. specifies the values of the  spins outside  $\Lambda_L$
as follows,  \be\label{abbc} \forall x\not\in \Lambda_L\;,\;\sigma(x):=
\cases{-1& if $x(2)\leq a\cdot L$, $x(1)<0$,\cr                  
       -1& if $x(2)\leq b\cdot L$, $x(1)\geq 0$,\cr
       +1& otherwise.\cr}
\ee
The $-$ b.c. specifies the values of the  spins outside $\Lambda_L$ 
as follows,
\be
\forall x\not\in \Lambda_L\;,\;\sigma(x):=-1\,.
\ee 
In $\Lambda_L$ we consider the Ising model defined by the Hamiltonian
\be\label{hamiltonian}
H_{\Lambda_L}=-\sum_{\bk{t,t'}\cap\Lambda_L\not=\emptyset}
J(t,t')\sigma(t)\sigma(t')\,,
\ee
where $\bk{t,t'}$ denotes a pair of nearest neighbours points
of the lattice $\bZ^2$, or the corresponding edge (considered as a
unit--length segment) with end--points
$t,t'$;  the coupling constants $J(t,t')$ are  given by 
\be\label{couplingconst}
J(t,t'):=\cases{1& if $t(2)\geq 0$ and $t'(2)\geq 0$,\cr
                h& otherwise, with $h> 0$.\cr}
\ee
Let $\beta$ be the inverse temperature. The Boltzmann factor is
$\exp\{-\beta H_{\Lambda_L}\}$
and the  Gibbs measures in $\Lambda_L$ with
$ab$ b.c., respectively $-$ b.c., are denoted by
\be
\bk{\,\cdot\,}_L^{ab}=\bk{\,\cdot\,}_L^{ab}(\beta,h)
\quad\hbox{resp.}\quad
\bk{\,\cdot\,}_L^{-}=
\bk{\,\cdot\,}_L^{-}(\beta,h)\,.
\ee
We introduce the dual lattice to $\bZ^2$ 
\be
\bZ^{2*}:=\{\,x=(x(1),x(2)):\,x+(1/2,1/2)\in\bZ^2\,\}\,,
\ee
and describe the spin configurations $\omega$ by a set of edges 
$\cE^{*}(\omega)$ of the dual lattice.
For each edge $e$ of $\bZ^2$ there is a unique edge $e^*$ of
$\bZ^{2*}$ which crosses $e$, which is written $e^*\dagger e$.
Let $\omega\in\{-1,+1\}^{\bZ^2}$ be a
spin configuration satisfying the $ab$ b.c.. We set
\be 
\cE^{*}(\omega):=\{e^*\subset\Lambda^*_L:\,\exists \,
\bk{t,t'}\dagger e^*
\;\hbox{with}\; \sigma(t)(\omega)\sigma(t')(\omega)=-1\,\}\,.
\ee 
We decompose the set $\cE^{*}(\omega)$ into connected components
and use  rule $A$ defined in the picture below in order to get
a set of disjoint simple lines called {\bf contours}.

\begin{picture}(400,100)(-275,-60)
\thicklines
\put(-105,0){\line(1,0){40}}
\put(-85,-20){\line(0,1){40}}
\put(-50,0){\vector(1,0){20}}
\put(-20,0){\line(1,0){15}}
\put(-5,-5){\oval(10,10)[tr]}
\put(0,-5){\line(0,-1){15}}
\put(0,20){\line(0,-1){15}}
\put(5,5){\oval(10,10)[bl]}
\put(5,0){\line(1,0){15}}

\put(-60,-30){\makebox(35,15)[bl]{ rule $A$}}
\end{picture}

Each  configuration $\omega$ satisfying the $ab$ b.c. is uniquely
specified  by a family $(\underline{\gamma}(\omega),\lambda(\omega))$
of disjoint contours;  all contours of $\underline{\gamma}=
\{\gamma_1,\gamma_2,\ldots\}$ are closed\footnote{
Let $A$ be a set of edges; the boundary $\delta A$ of $A$ is the set
of $x\in\bZ^{2*}$ such that there is an odd number of edges of $A$
adjacent to $x$. $A$ is  {\bf closed} if $\delta A=\emptyset$ and
{\bf open} if $\delta A\not=\emptyset$.}
and  $\lambda$ is open, 
with end--points $t_l^L$ and $t_r^L$. We call $\lambda$ the 
{\bf phase separation line}\footnote{ 
As already mentioned in the introduction we make a 
distinction between the concept of ``phase separation line", which 
is defined for each configuration at the scale of the lattice 
spacing, and the concept of ``interface", which is associated with 
the fact that there is a separation of the two phases in the model 
due to our choice of boundary condition. The ``interface" concept
emerges at a scale large enough so that it is a non--random object,
whose free energy is given in terms of the surface tension.}.
Conversely, a family of contours 
$(\underline{\gamma}',\lambda')$ is called 
{\bf $ab$ compatible}\footnote{
 To be precise we should say that
$(\underline{\gamma}',\lambda')$ is $ab$ compatible in $\Lambda_L$.
Compare this notion of compatibility with the notion
of compatibility  used in the high--temperature expansion 
(see subsection \ref{subduality}).}
if there exists $\omega$ such that $\omega$
satisfies the $ab$ b.c. and 
$\underline{\gamma}(\omega)=\underline{\gamma}'$,
$\lambda(\omega)=\lambda'$. In the same way
each  configuration $\omega$ satisfying the $-$ b.c. is uniquely 
specified by a family $\underline{\gamma}$ of closed contours and we
have a notion of $-$ compatibility.

For each contour $\eta$, closed or open, we define a set of edges
of $\bZ^2$,
\be
{\rm co}(\eta):=
\{\,e\in \bZ^2:\,\exists e^*\in\eta,\;e\dagger e^*\,\}\,.
\ee
The Boltzmann weight of $\eta$ is
\be
w(\eta):=\prod_{\bk{t,t'}\in{\rm co}(\eta)}       
\exp(-2\beta J(t,t'))\,.
\ee
Next we define three (normalized) partition
functions, $Z^{ab}(\Lambda_L)$, $Z^{ab}(\Lambda_L|\lambda)$ and
$Z^{-}(\Lambda_L)$. By definition
\be\label{2.14}
Z^{ab}(\Lambda_L):=\sum_{ \omega\;{\rm with}\; ab\;b.c.}
w(\lambda(\omega))
\prod_{\gamma\in\underline{\gamma}(\omega)}w(\gamma)\,;
\ee
\be\label{2.15}
Z^{ab}(\Lambda_L|\lambda):=\sum_{ {\sc \omega\;{\rm with}\; ab\;b.c.:
\atop \sc \lambda(\omega)=\lambda}}
\prod_{\gamma\in\underline{\gamma}(\omega)}w(\gamma)\,;
\ee
\be\label{2.16}
Z^{-}(\Lambda_L):=\sum_{ \omega\;{\rm with}\; -\,b.c.}
\prod_{\gamma\in\underline{\gamma}(\omega)}w(\gamma)\,.
\ee
We  define a weight $q_L(\lambda)=
q_L(\lambda;\beta,h)$ for each phase separation line $\lambda$ of a
configuration $\omega$  satisfying
the $ab$ b.c.,
\be\label{qweight}
q_L(\lambda):=\cases{w(\lambda){\displaystyle Z^{ab}(\Lambda_L|\lambda)\over 
\displaystyle Z^{-}(\Lambda_L)}& if $\lambda$ is $ab$ compatible in $\Lambda_L$,\cr
0& otherwise.\cr}
\ee

\subsection{Surface tension}

Consider the model defined in $\Lambda_L$, with coupling constants
$J(t,t')\equiv 1$, i.e. $h=1$ in (\ref{couplingconst}).  For each 
$\omega$ compatible with the $ab$ b.c. there is a well-defined phase
separation line $\lambda(\omega)$ with end--points $t_l^L$ and 
$t_r^L$. Let $n=(n(1),n(2))$ be the unit vector in $\bR^2$ which is 
perpendicular to the straight line passing through $t_l^L$ and
$t_r^L$.  By definition  {\bf the surface tension} $\stau(n;\beta)=
\stau(n)$ is
\be
\stau(n)=\stau(n(1),n(2)):=
-\lim_{L\ra\infty}{1\over \|t_l^L-t_r^L\|}
\log{Z^{ab}(\Lambda_L)\over Z^{-}(\Lambda_L)}\,,
\ee 
where $\|t_l^L-t_r^L\|$ is the Euclidean distance between
$t_l^L$ and $t_r^L$. By symmetry of the model we have
\be
\stau(n(1),n(2))=\stau(-n(1),-n(2))=\stau(n(2),-n(1))
=\stau(n(2),n(1))\,.
\ee
Using (\ref{qweight}) we can write
\be
\stau(n)=
-\lim_{L\ra\infty}{1\over \|t_l^L-t_r^L\|}\log
\sum_{{\sc \lambda(\omega):\atop\sc\omega\;{\rm with}\; ab-b.c.}}
q_L(\lambda(\omega))\,.
\ee
We extend the definition of the surface tension
to $\bR^2$ by homogeneity,
\be
\stau(x):=\|x\|\stau(x/\|x\|)\,.
\ee

\begin{pro}\label{prosurface}
The surface tension is a uniformly Lipschitz convex function on 
$\bR^2$, such that $\stau(x)= \stau(-x)$.
It is identically zero above the critical
temperature and strictly positive for all $x\not =0$ when the
temperature is  strictly smaller than the critical temperature.

The main property of $\stau$ is the sharp triangle inequality. For
all  $\beta>\beta_c$ there exists a strictly positive constant 
$\Delta=\Delta(\beta)$
such that for any $x$, $y\in\bR^2$, the norm $\stau(\,\cdot\,)$
satisfies
\be\label{striangle}
\stau(x)+\stau(y)-\stau(x+y)\geq
\Delta(\|x\|+\|y\|-\|x+y\|)\,. 
\ee
Let $x(\theta):=(\cos \theta,\sin\theta)$ and $\stau(\theta):=
\stau(x(\theta))$. Then the best constant $\Delta$ is 
\be\label{stiffness}
\Delta:=\inf_{\theta}\left({d^2\over
d\theta^2}\stau(\theta)+\stau(\theta)\right)>0\,. 
\ee 
\end{pro} 

{\bf Remark:} 
The first part of the proposition follows from \cite{MW} or
\cite{LP}. The second part is proven in  section \ref{sti}.
(\ref{striangle}) was introduced and proven by Ioffe in  \cite{I}.
The statement of (\ref{striangle}) is different in 
\cite{I}, but equivalent to the present one (see proof of 
Proposition \ref{stiequivalence}). 
The constant $\Delta$ is not optimal in \cite{I}.
(\ref{stiffness}) is  called the {\bf positive stiffness
property}. Geometrically, (\ref{stiffness})  means that the 
curvature of  the Wulff shape is bounded above
by $1/\Delta$.

\subsection{Duality}\label{subduality}

A basic property of the 2D Ising model is self-duality. As a
consequence of that property many questions about the model below
the critical temperature can be translated into dual questions for
the dual model above the critical temperature. For example, questions
about the surface tension are translated into questions about the
correlation length. We refer to 
\cite{PV1} for a more complete discussion and recall here the main
results, which we shall use in section \ref{Ellipse}.

Consider the model defined in the box $\Lambda_L$ with coupling
constants given by (\ref{couplingconst}). We suppose that we
have $-$ b.c.. Let 
\be
\cE_L:=\{\,\bk{t,t'}:\,t\;\hbox{or}\;
t'\in\Lambda_L\,\}\,,
\ee
be the set of edges in the sum (\ref{hamiltonian}). The dual set
of edges is
\be
\cE^*_L:=\{\,e^*:\,\exists 
e\in\cE_L\,,\,
e^*\dagger e\,\}\,,
\ee
and the dual model is defined on the dual box
\bea\label{lambda*}
\Lambda_L^*&:=&\{x\in\bZ^{2*}:\,\exists e^*\in
\cE^*_L\,,\, x\in e^*\,\}\\
&=&\{\,x\in\bZ^{2*}:\,
|x(1)|\leq (L+1)/2\;,\;-1/2\leq x(2)\leq L+1/2\,\}\nonumber
\,.
\eea
The dual Hamiltonian is the free boundary
condition (free b.c.) Hamiltonian, that is,  
\be\label{dualhamiltonian}
H_{\Lambda_L^*}=-\sum_{\bk{t,t'}\subset\Lambda_L^*}J^*(t,t')
\sigma(t)\sigma(t')\,.
\ee
In (\ref{dualhamiltonian}) the dual coupling constants are
related to the coupling constants (\ref{couplingconst}) as
follows:
\be\label{dualcoupling}
J^*(t,t'):=\cases{h^* & if $t(2)=t'(2)=-1/2$,\cr
 1& otherwise.\cr}
\ee
$h^*$ and the dual temperature $\beta^*$ are defined by
\bea
\tanh\beta^*&:=&\exp\{-2\beta\}\,;\\
\tanh\beta^*h^*&:=&\exp\{-2\beta h\}\,.
\eea
The critical inverse temperature $\beta_c$ is characterized by
\be
\tanh\beta_c:=\exp\{-2\beta_c\}\,.
\ee
Notice that when $h$ or $\beta$ are small, then $h^*$ or $\beta^*$
are large. The Gibbs measure in $\Lambda_L^*$ with free boundary
condition is denoted by 
$\bk{\,\cdot\,}_{\Lambda_L^*}=
\bk{\,\cdot\,}_{\Lambda_L^*}(\beta^*,h^*)$. In this paper 
$\beta>\beta_c$ so that $\beta^*<\beta_c$. For those values of
$\beta^*$ there is a unique Gibbs measure on $\bZ^2$, which
we denote by $\bk{\,\cdot\,}=
\bk{\,\cdot\,}(\beta^*)$\footnote{
This measure is the limit of Gibbs measures in finite
subsets $\Lambda$ with free boundary condition, when
$\Lambda\uparrow\bZ^2$. As long as $\beta^*<\beta_c$, the choice of
the boundary conditions does not matter since there is a unique
Gibbs state. Notice that $\bk{\,\cdot\,}(\beta^*)$ is not the limit of
the measures $\bk{\,\cdot\,}_{\Lambda_L^*}(\beta^*,h^*)$ when
$L\ra\infty$, since the subsets $\Lambda_L^*$ do not converge to
$\bZ^2$. There is a limiting measure for 
$\bk{\,\cdot\,}_{\Lambda_L^*}$ when $L\ra\infty$, which is defined on
the semi--infinite lattice $\bL^*:=\{x\in\bZ^{2*}:\,
x(i)\geq-1/2\,,i=1,2\}$, and which depends on $\beta^*$ {\it and}
$h^*$ \cite{FP2}.}. The most important quantity for
the 2D Ising model with free b.c. is the covariance function, or
two--point function, 
\be
\bk{\sigma(t)\sigma(t')}(\beta^*)\,.
\ee
The {\bf decay-rate} of the covariance function is defined for
all $t,t'\in\bZ^{2*}$ as
\be
\tau(t-t')=
\tau(t-t';\beta^*)
:=-\lim_{{\sc n\in\bN\atop\sc n\ra\infty}}
{1\over n}\log
\bk{\sigma(nt)\sigma(nt')}(\beta^*)\,.
\ee
 
\begin{pro}\label{produality}
For the 2D Ising model the surface tension $\stau(x;\beta)$
and the decay--rate $\tau(x;\beta^*)$ are equal,
\be
\stau(x;\beta)=\tau(x;\beta^*)\quad\forall x\,.
\ee
\end{pro}
For a proof see \cite{BLP2}.

Proposition \ref{produality} indicates that the decay--rate 
and surface tension are dual quantities. Moreover,
properties of the phase separation line $\lambda$
at $\beta$ are related to properties of the covariance function at
$\beta^*$ through the random--line representation of the covariance
(see \cite{PV1} for detailed discussion). The random-line
representation  follows from the high--temperature
expansion. The terms of this expansion are indexed by sets of edges,
called contours. Throughout the paper we use the following
notations:  if $A\subset\bZ^{2*}$, then $\cE^*(A)$ is the set of all
edges of $\bZ^{2*}$ with both end--points in $A$.
Consider the partition
function in $\Lambda_L^*$ with free b.c., which can be written as
\be
\sum_{\sigma(t),\,t\in\Lambda_L^*}\prod_{\bk{t,t'}\subset\Lambda_L^*}
  \cosh J^*(t,t')(1+\sigma(t)\sigma(t')\tanh J^*(t,t'))\,.
\ee
We expand the product; each term of the expansion is labeled by a set
of  edges $\bk{t,t'}$: we specify the edges corresponding to factors
$\tanh J^*(t,t')$. Then we sum over $\sigma(t)$, $t\in\Lambda$; after
summation only terms labeled by sets of edges of the dual lattice
$\bZ^{2*}$ with empty boundary
give a non--zero contribution. We decompose this set uniquely into 
a family of connected closed contours using the rule $A$. Any such
family of contours is called {\bf compatible}\footnote{
To be precise we should say compatible in $\Lambda_L^*$, since each
contour is a subset $\cE^*(\Lambda_L^*)$.
 A family of closed contours in $\Lambda_L^*$ is
compatible  if and only if they are disjoint according to rule $A$.
This is a purely geometrical property, contrary to the definition of
$-$ compatibility. A family of closed contours which is $-$ compatible
in $\Lambda_L$ is also compatible in $\Lambda_L^*$. Because of our
choice of $\Lambda_L$ the converse is also true, but in general
$-$ compatibility does not imply compatibility.}.
For each (closed) contour $\gamma$ we set 
\be
w^*(\gamma):=\prod_{\bk{t,t'}\in\gamma}\tanh J^*(t,t')\,,
\ee
and we introduce a normalized partition function
\be
Z(\Lambda_L^*):=\sum_{{\sc\underline{\gamma}:\atop
\sc {\rm comp.
\,in}\,\Lambda_L^*}}\prod_{\gamma\in\underline{\gamma}} w^*(\gamma)\,.
\ee
We treat the numerator of the two--point function 
$\bk{\sigma(t)\sigma(t')}_{\Lambda_L^*}$ in a similar way. In this
case all non--zero terms of the expansion are labeled by compatible
families $(\underline{\gamma},\lambda)$, where all $\gamma\in
\underline{\gamma}$ are closed, $\lambda$ is open with end--points
$t,t'$. Given an open contour $\lambda$, we introduce a partition
function as in (\ref{2.15}), 
\be
Z(\Lambda_L^*|\lambda):=
\sum_{ {\sc \underline{\gamma}:
\atop \sc (\underline{\gamma},\lambda)\,{\rm comp.}}}
\prod_{\gamma\in\underline{\gamma}} w^*(\gamma)\,.
\ee
The next two formulas are fundamental. For each open contour
$\lambda$ we define the weight of the contour as
\be\label{weightdef}
q^*_L(\lambda)=q^*_L(\lambda;\beta^*,h^*):=
\cases{w^*(\lambda)
{\displaystyle Z(\Lambda_L^*|\lambda)\over \displaystyle Z(\Lambda_L^*)}&if 
$\lambda\subset\cE^*(\Lambda_L^*)$,\cr
0& otherwise.\cr} 
\ee 
Using this weight we get a random--line representation for the
two--point function $\bk{\sigma(t)\sigma(t')}_{\Lambda_L^*}$ as 
\be\label{randomrepresentation}
\bk{\sigma(t)\sigma(t')}_{\Lambda_L^*}=
\sum_{\lambda:t\ra t'}q^*_L(\lambda)\,.
\ee

There are  similar representations for $\bk{\sigma(t)\sigma(t')}$
and for
\be
\lim_{L\ra\infty}
\bk{\sigma(t)\sigma(t')}_{\Lambda_L^*}(\beta^*,h^*)
=\bk{\sigma(t)\sigma(t')}_{\bL^*}(\beta^*,h^*)\, 
\ee
when $\beta^*<\beta_c$. 

Let $\lambda$ be such that $\delta\lambda=\{t,t'\}$; we also write
$\lambda:t\ra t'$. 
Given $\lambda:t\ra t'$ we can define weights
$q^*(\lambda;\beta^*)$ and $q^*_{\bL^*}(\lambda;\beta^*,h^*)$ such 
that
\be
\bk{\sigma(t)\sigma(t')}(\beta^*)=
\sum_{\lambda:t\ra t'}q^*(\lambda;\beta^*)\,,
\ee
and
\be
\bk{\sigma(t)\sigma(t')}_{\bL^*}(\beta^*,h^*)=
\sum_{\lambda:t\ra t'}
q^*_{\bL^*}(\lambda;\beta^*,h^*)\,. 
\ee

\begin{pro}\label{prorandom}
Let $\beta^*<\beta_c$. Let $\lambda_1$ and $\lambda_2$ be two
open contours such that $\lambda:=\lambda_1\cup\lambda_2$ is an
open contour and $\lambda\subset\cE(\Lambda_L^*)$. 
Then
\be
q^*_L(\lambda;\beta^*,h^*)\geq q^*_{\bL^*}(\lambda;\beta^*,h^*)
\quad \hbox{and}\quad
q^*_L(\lambda)\geq q^*_L(\lambda_1) q^*_L(\lambda_2)\,.
\ee
If $\delta\lambda=\{t_l^L,t_r^L\}$,
then
$q_L(\lambda;\beta^*,h^*)$ is equal to (\ref{qweight}), that is,
\be
q^*_L(\lambda;\beta^*,h^*)=q_L(\lambda;\beta,h)\,.
\ee

\end{pro}

Proposition \ref{prorandom}\footnote{Notice that Proposition 
\ref{prorandom} does not
imply Proposition \ref{produality}, see  discussion in
subsection \ref{correlationlength}} is a 
key--result which is proven in
\cite{PV1}. It allows us to study
properties of the phase
separation line through the two--point correlation function 
$\bk{\sigma(t_l^L)\sigma(t_r^L)}_{\Lambda_L^*}(\beta^*,h^*)$. From
Proposition \ref{prorandom} and GKS inequalities we get the
interesting inequalities \bea
\sum_{{\sc\lambda:\delta\lambda=\{t,t'\}\atop
\sc\lambda\subset\cE^*(\Lambda_L^*)}}q^*_{\bL^*}(\lambda)
&\leq&
\sum_{{\sc\lambda:\delta\lambda=\{t,t'\}\atop
\sc\lambda\subset\cE^*(\Lambda_L^*)}}q^*_L(\lambda)
=\bk{\sigma(t)\sigma(t')}_{\Lambda_L^*} \\
&\leq&
\bk{\sigma(t)\sigma(t')}_{\bL^*}=
\sum_{{\sc\lambda:\delta\lambda=\{t,t'\}\atop
\sc\lambda\subset\cE^*(\bL^*)}}q^*_{\bL^*}(\lambda)\,.\nnm
\eea

\subsection{Wall free energy}\label{subfreewall}

The last thermodynamical quantity, which enters into the 
description of
the properties of the interface, is the wall free energy. We define the
difference of the contributions of the wall to the free energy when
the bulk phase is the $+$ phase, respectively the $-$ phase, 
as\footnote{
The definition of $\staubd$ differs from the
analogous quantity used in \cite{PV1} or \cite{PV2}, because in these
papers the reference bulk phase is the  $+$ phase and here it is the
$-$ phase.}
\be
\staubd=\staubd(\beta,h):=-\lim_{L\ra\infty}{1\over 2L+1}\log
{Z^{00}(\Lambda_L)\over Z^{-}(\Lambda_L)}
\,,
\ee
where $Z^{00}(\Lambda_L)$ is the partition function with 
$a=b=0$. There is a proposition analogous to Proposition
\ref{produality}, which relates $\staubd$ to the decay--rate of the
boundary two--point function of the dual model (see \cite{PV1})
 
\begin{pro}\label{produalitybd}

Let $\beta>\beta_c$. Let $t,t'\in\bL^*$, $t(2)=t'(2)=-1/2$. Then
\be
-\lim_{n\ra\infty}{1\over n}\log
\bk{\sigma(nt)\sigma(nt')}_{\bL^*}(\beta^*,h^*)=
\|t-t'\|\cdot\staubd(\beta,h)\,.
\ee
\end{pro}

The quantity $\staubd(\beta,h)$ allows to detect the wetting
transition through Cahn's criterium
(see \cite{FP1} and \cite{FP2}). Since $h>0$, 
$0<\staubd(\beta,h)\leq \stau(\beta)$. There is partial wetting of
the wall if and only if  $\staubd(\beta,h)< \stau(\beta)$; this occurs
if and only if $h<h_w$, where $h_w$ has been computed by Abraham
\cite{A}. The transition value $h_w(\beta)$ is the solution of the
equation 
\be 
\exp\{2\beta\} \{\cosh 2\beta-\cosh 2\beta h_w(\beta)\}= 
\sinh 2\beta\,. 
\ee 
By duality we show in subsection  \ref{correlationlength} that we get
finite size effects for the two--point function when
$h^*>h_c^*$, where 
$h_c^*(\beta^*):=(h_w(\beta))^*$\footnote{ In this paper we always
assume that $0<h<\infty$, so that we also have $0<h^*<\infty$. Notice
that $0<h_w(\beta)<1$ for any $0<\beta<\beta_c$; consequently
$1<h_c^*(\beta^*)<\infty$ for any $\beta_c<\beta^*<\infty$.}.

\begin{figure}[h]
 \centerline{\psfig{figure=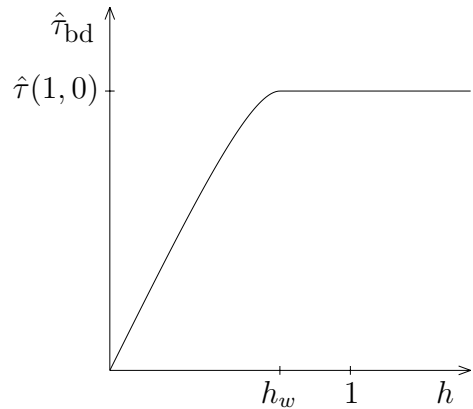,height=50mm}}
  \figtext{
 	\writefig	8.1	0.20	{$1$}
 	\writefig	7.0	0.20	{$h_w$}
 	\writefig	9.33	0.20	{$h$}
 	\writefig	4.20	5.1	{$\hat{\tau}_{\mbox{\rm \footnotesize bd}}$}
 	\writefig	3.7	4.25	{$\hat{\tau}(1,0)$}
  }
 \caption[]
{$\staubd$ as a function of the magnetic field $h$, for $\beta=1.4\beta_c$.}
 \label{fig_epspath}
\end{figure}
 

\section{The variational problem}\label{Variational}
\setcounter{equation}{0}

The interface is a macroscopic
non--random object, whose properties are described by a functional
involving the surface tension. In $Q$ the interface is a
simple rectifiable curve $\cC$ with end--points $A:=(-1/2,a)$,
$0<a<1$, and
$B:=(1/2,b)$, $0<b<1$. We denote by $w_Q=\{\,x\in Q: \,x(2)=0\,\}$
the wall and by $|\cC\cap w_Q|$ the length of the portion of the
interface in contact with the wall $w_Q$.
Suppose that $[0,w]\ra Q$, $s\mapsto \cC(s)=(u(s),v(s))$,
is a parameterization of the interface. The free energy of the 
interface $\cC$ can be written
\be
\W(\cC):=\int_0^w\stau(\dot{u}(s),\dot{v}(s))ds
+|\cC\cap w_Q|\cdot
\Big[\staubd-\stau(1,0)\Big]\,,
\ee
(because the function $\stau(x(1),x(2))$ is positively homogeneous 
and $\stau(x(1),x(2))=\stau(-x(2),x(1))$).
The interface at equilibrium is the
minimum of this functional. Therefore we have to solve 
the 

{\bf Variational problem}: {\sl Find the minimum of the functional 
$\W$ among all simple rectifiable open curves in $Q$ with extremities
$A=(-1/2,a)$ and $B=(1/2,b)$.}

Let $\cD$ be the straight line from $A$ to $B$ and $\cW$ be
the curve composed of the  following three straight line segments: 
from $A$ to a point $w_1$  on the wall, then along the wall from $w_1$
to $w_2$, and finally from $w_2$ to $B$. The points $w_1$ and $w_2$
are such that the angles between the first segment and the wall and
between the last segment and the wall are equal,
chosen\footnote{ This choice leads to a different sign at the 
right--hand side of the Herring-Young equation (\ref{Young}) than
in \cite{PV2} formulae (1.5) or (4.60); in these latter references
we use  $\pi-\theta$ instead of $\theta$.} in
the interval $[0,\pi/2]$, and solutions of 
\be
\cos\theta_Y \, \tau(\theta_Y) - \sin\theta_Y \, \tau'(\theta_Y) =
\staubd\,,\label{Young}
\ee
which is known as the Herring-Young equation. 
(For the case under consideration the existence of $\theta_Y$ is an 
immediate consequence of the Winterbottom construction).

\begin{pro}
Let $\theta_Y$ be the solution of the Herring-Young equation 
\req{Young}.
\begin{enumerate}
\item
If $\tan\theta_Y\leq a+b$ then the minimum of the
variational problem is given by the curve $\cD$.
\item
If $\pi/2>\theta_Y>\arctan(a+b)$ then the minimum of the variational 
problem is given by $\cD$ if $\W(\cD)<\W(\cW)$, by $\cW$ if
$\W(\cD)>\W(\cW)$ and by both $\cD$ and $\cW$ if $\W(\cD)=\W(\cW)$.
\end{enumerate}
\end{pro}
\prf
The proof is an easy consequence of the two following lemmas. 
Lemma \ref{lem3.1} states that the minimum is a polygonal line.

\begin{lem}\label{lem3.1}
Let $\cC$ be some simple rectifiable  parameterized curve with initial 
point $A$ and final point $B$.

If $\cC$ does not intersect the wall, then
\be
\W(\cC)\geq \W(\cD)
\ee
with equality if and only if $\cC$=$\cD$.

If $\cC$ intersects the wall,  
let $t_1$  be the first time $\cC$ touches  the wall and $t_2$ the 
last time  $\cC$ touches  the wall. Let $\widehat{\cC}$ be the curve 
given by  three segments from $A$ to $\cC(t_1)$, from $\cC(t_1)$ to 
$\cC(t_2)$ and from $\cC(t_2)$ to $B$. Then
\be
\W(\cC)\geq \W(\widehat{\cC})\,.
\ee
Equality holds if and only if $\cC=\widehat{\cC}$.
\end{lem}

\prf
Since $\stau$ is convex and homogeneous, we have in the first case
by  Jensen's inequality
\be
\W(\cC) =w{1\over w}\int_0^w\stau(\dot{u}(s),\dot{v}(s))ds \geq
w\stau({1\over w}\int_0^w\dot{u}(s)ds,{1\over w}\int_0^w\dot{v}(s)ds)
 =\W(\cD)\,.
\ee
The inequality is strict if $\cC\neq\cD$ as is easily seen using the 
sharp triangle inequality \req{striangle}.

In the second case we apply Jensen's inequality to the part of $\cC$
between $A$ and $\cC(t_1)$ and between $\cC(t_2)$ and $B$ to compare 
with the corresponding straight segments of $\widehat{\cC}$.
Combining Jensen's inequality and the fact that $\staubd\leq\stau$,
we can also compare the part of $\cC$ between $\cC(t_1)$ and
$\cC(t_2)$ with  the corresponding straight  segment of
$\widehat{\cC}$. \qed

\begin{lem}\label{lem3.2}
Let $\widehat{\cC}$ be a polygonal line from $A$ to 
$\hat{w}_1\in w_Q$, then from $\hat{w}_1$ to
$\hat{w}_2\in w_Q$, and finally from $\hat{w}_2$ to $B$.
Let $\theta_Y$ be the solution of the Herring-Young equation 
\req{Young}.\\
If $\pi/2>\theta_Y > \arctan(a+b)$ then
\be\label{wetting}
\W(\widehat{\cC})\geq \W(\cW)\,,
\ee
with equality if and only if $\widehat{\cC}=\cW$.\\
If $\arctan(a+b)\geq\theta_Y$
\be
\W(\widehat{\cC})> \W(\cD)\,.
\ee
\end{lem}
\prf

Let $\theta_1\in(0,\pi/2)$ be the angle of the straight segment
of  $\widehat{\cC}$, from $A$ to $\hat{w}_1$, with the wall $w_Q$, and
$\theta_2\in(0,\pi/2)$ be the angle of the straight segment
of  $\widehat{\cC}$, from $\hat{w}_2$ to $B$, with the wall $w_Q$.
We have
\bea
\W(\widehat{\cC})&=&\tau(\theta_1){a \over \sin\theta_1} + 
\staubd(1-{a\over \tan\theta_1}-{b\over\tan\theta_2}) + 
\tau(\theta_2){b \over \sin\theta_2}\\
&=& g(\theta_1,a)+g(\theta_2,b)\,,\nonumber
\eea
where we have introduced
\be\label{definitionofg}
g(\theta,x) := \tau(\theta){x \over \sin\theta} +
\staubd(1/2-{x\over\tan\theta})\,.
\ee
Let $\theta_Y$ be defined as the solution of the Herring-Young 
equation \req{Young}, so that
\be
{\partial\over
\partial\theta}g(\theta_Y,x)={x\over \sin^2\theta_Y}
( \sin\theta_Y \, \tau'(\theta_Y) -\cos\theta_Y \, \tau(\theta_Y) + 
\staubd)=0
\,. 
\ee
The second derivative of $g(\theta,x)$ is
\be\label{secondderivative}
{\partial^2 \over
\partial\theta^2}g(\theta,x)
=
{x(\tau(\theta)+\tau''(\theta)) \over
\sin\theta}-{2\over\tan\theta}{\partial\over
\partial\theta}g(\theta,x)\,.
\ee
Therefore, for $\theta\in (0,\pi/2)$, we have
\be\label{3.122}
{\partial\over\partial\theta}g(\theta,x)=
x\int_{\theta_Y}^{\theta}
\exp\{-\int_{\gamma}^{\theta}{2\over \tan\alpha}\, d\alpha\}
\,{\tau(\gamma)+\tau''(\gamma) \over\sin\gamma}\,d\gamma\,.
\ee
Since $\tau$ has positive stiffness, i.e. $\tau(\theta)+
\tau''(\theta)>0$, (\ref{3.122}) implies that $\theta_Y$ is an
absolute  minimum of $g(\theta,x)$ over the interval $(0,\pi/2)$, and that $g$
is 
strictly monotonous over the intervals $(\theta_Y,\pi/2)$ and
$(0,\theta_Y)$.

A necessary and sufficient condition, that we can construct a simple
polygonal line $\widehat{\cC}$ as above, is
\be
{a\over\tan\theta_1}+{b\over\tan\theta_2}\leq 1\,.
\ee
In particular $\theta_1\in [\theta_a,\pi/2]$ where
$\theta_a:=\arctan a$, and $\theta_2\in [\theta_b,\pi/2]$ where
$\theta_b:=\arctan b$. Similarly
$\cW$ is a simple curve in $Q$ if and only if
\be\label{3.12}
\theta_Y\in [\arctan a+b,\pi/2)\,.
\ee
From the preceding results we have
\be\label{3.11}
\W(\widehat{\cC})\geq g(\theta_1^*,a)+g(\theta_2^*,b)\,,
\ee
with 
\bea
\theta_1^*&=&\cases{\theta_Y & if $\theta_Y\in [\theta_a,\pi/2]$,\cr
\theta_a & otherwise,\cr}\\
\theta_2^*&=&\cases{\theta_Y & if $\theta_Y\in [\theta_b,\pi/2]$,\cr
\theta_b & otherwise.\cr}\nnm
\eea
If (\ref{3.12}) holds, then (\ref{3.11}) implies 
$\W(\widehat{\cC})\geq \W(\cW)$. 
If (\ref{3.12}) does not hold,
then the two segments from $A$ to the wall,
and from $B$ to the wall intersect at some point
$P$. Let $\widehat{\cW}$ be the simple polygonal curve going from $A$
to $P$, then from $P$ to $B$. We have (this follows from
Lemma \ref{lem3.1} and $\stau(1,0)\geq\staubd$)
\be
g(\theta_Y,a)+g(\theta_Y,b)\geq \W(\widehat{\cW})\,.
\ee
Applying again Lemma \ref{lem3.1} we get
\be
\W(\widehat{\cW})> \W(\cD)\,.
\ee
\qed

\begin{figure}
 \centerline{\psfig{figure=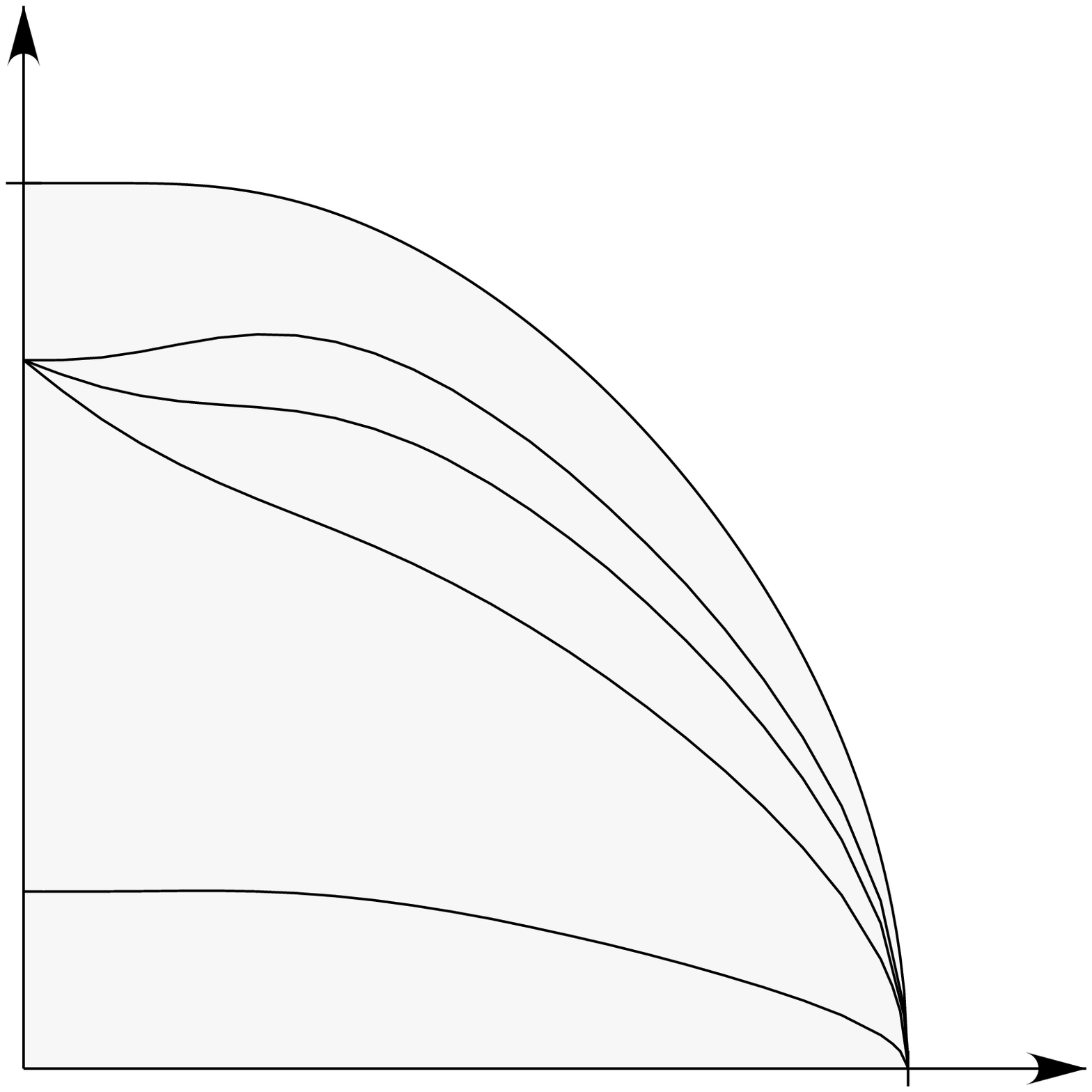,height=60mm}}
  \figtext{
 	\writefig	4.2	6.2	{\footnotesize $h$}
 	\writefig	4.1	4.49	{\footnotesize $0.8$}
 	\writefig	4.1	1.63	{\footnotesize $0.2$}
 	\writefig	9.2	0.4	{\footnotesize $T_c$}
 	\writefig	9.97	0.4	{\footnotesize $T$}
 	\writefig	6.3	4.7	{\footnotesize $i$}
 	\writefig	6.3	4.3	{\footnotesize $ii$}
 	\writefig	6.3	3.68	{\footnotesize $iii$}
 	\writefig	6.3	1.74	{\footnotesize $iv$}
  }
 \caption[]{A sequence of phase coexistence lines, separating the phase in
 which the interface is a straight line and the phase in which it is pinned
 to the wall, for different values of the parameters $a$ and $b$. For each of
 the four curves, for all values of the parameters $\beta$ and $h$ above the
 curve, the interface is the straight line and for all values of the parameters
 below the curve the interface is pinned to the wall. The shaded area
 correspond to the value of $(\beta,h)$ so that
 $\staubd(\beta,h)<\stau((1,0);\beta)$. The four curves correspond to: i)
 $a=0.1$, $b=0.1$; ii) $a=0.1$, $b=0.2$; iii) $a=0.1$, $b=0.4$; iv) $a=0.4$,
 $b=0.4$. Observe that the system in case 1) exhibits reentrance: if we fix the
 value of the magnetic field near $0.8$ and increase the temperature from 0 to
 $T_c$, the system changes from phase I to phase II and then to phase I again
 (see also Fig. \ref{fig_zoom}).}
 
 \label{fig_phdiag}
\end{figure}
\begin{figure}
 \centerline{\psfig{figure=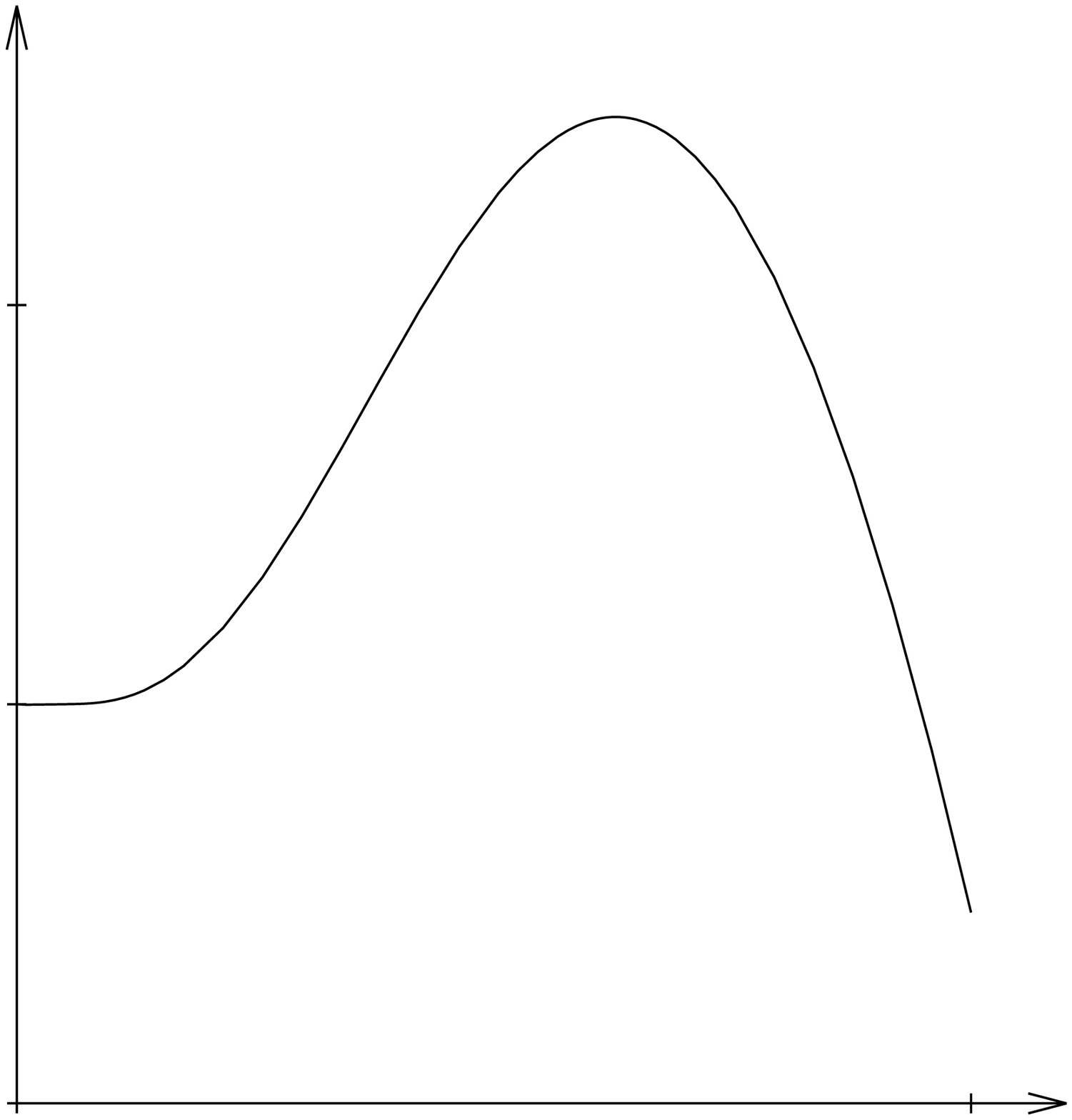,height=50mm}\hspace{1cm}
 \psfig{figure=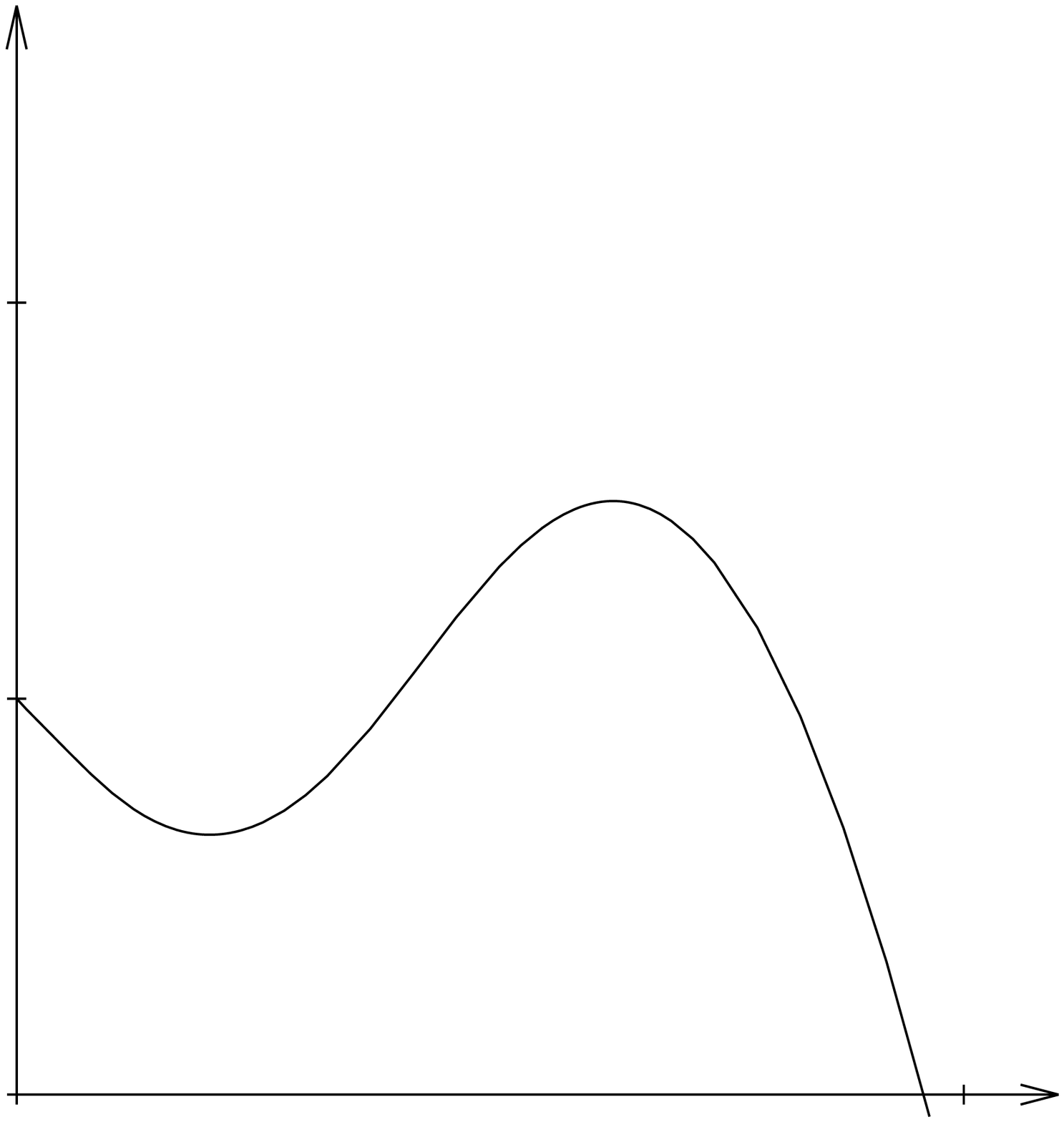,height=50mm}}
  \figtext{
 	\writefig	1.27	2.25	{\footnotesize $0.80$}
 	\writefig	1.27	0.53	{\footnotesize $0.78$}
 	\writefig	1.27	4.05	{\footnotesize $0.82$}
 	\writefig	6.27	0.3	{\footnotesize $1$}
 	\writefig	6.67	0.3	{\footnotesize $T$}
 	\writefig	1.72	5.1	{\footnotesize $h$}
 	\writefig	7.2	2.25	{\footnotesize $0.80$}
 	\writefig	7.2	0.53	{\footnotesize $0.78$}
 	\writefig	7.2	4.05	{\footnotesize $0.82$}
 	\writefig	12.2	0.3	{\footnotesize $1$}
 	\writefig	12.6	0.3	{\footnotesize $T$}
 	\writefig	7.65	5.1	{\footnotesize $h$}
  }
 \caption[]{This figure shows part of the phase coexistence line for $a=0.1,
 b=0.1$ (left), and $a=0.1,b=0.12$ (right). For values of the parameters
 $\beta$ and $h$ below these curves the interface is pinned, while it is a
 straight line above these curves. In case 2) the system has even one more
 transition in temperature for $h$ slightly smaller than $0.8$.
 }
 \label{fig_zoom}
\end{figure}


\section{Concentration properties}\label{Ellipse}
\setcounter{equation}{0}

By duality, properties of interfaces at temperatures below $T_c$ are 
related to properties of the random--line representation of the
two--point function of the Ising model at temperatures above $T_c$.
The results of this section are
given for the two--point function and are valid for all
temperatures above the critical temperature $T_c$. Our concentration
results are based on the sharp triangle inequality, which allows us
to improve  Propositions 6.1 and 6.2  of \cite{PV1}. The
results are essentially optimal.

The random--line representation for the
two--point function $\bk{\sigma(t)\sigma(t')}_{\Lambda_L^*}$ is
the formula
\be
\bk{\sigma(t)\sigma(t')}_{\Lambda_L^*}=
\sum_{\lambda:t\ra t'}q^*_L(\lambda)\,.
\ee
On the set of all open contours $\lambda$, such that 
$\delta\lambda=\{t,t'\}$, $q^*_L(\lambda)$ defines a measure
whose total mass is $\bk{\sigma(t)\sigma(t')}_{\Lambda_L^*}$.
The same is true for the similar representations of
$\bk{\sigma(t)\sigma(t')}(\beta^*)$ or
$\bk{\sigma(t)\sigma(t')}_{\bL^*}$.
It is therefore important to have good upper and lower bounds for
these quantities. We recall some basic results. Proposition 
\ref{lowerboundpro} is proven in \cite{MW} and the last part 
in \cite{PV1}; Proposition \ref{propv} is proven in \cite{PV1}.

We set $\Sigma^*:=\{x\in \bL^{*}:\,x(2)=-1/2\,\}$ and $\Sigma^*_L := \Sigma^* \cap
\Lambda^*_L$.
\begin{pro}\label{lowerboundpro}
Let $\beta^*<\beta_c$.
\begin{enumerate}
\item
There exists  $K$ such that
\footnote{We need the bound
(\ref{lowerbound}) for all temperatures $\beta^*<\beta_c$. Such
results have been obtained perturbatively in
\cite{Pl}, \cite{BF} and \cite{DKS} for example.
These lower bounds are essential in the proof of Proposition
\ref{lowerbd}.}
\be\label{lowerbound}
\bk{\sigma(x)\sigma(y)}\geq K{\exp\{-\stau(y-x)\}\over
\|x-y\|^{1/2}}\,.
\ee
\item
Let  $\stau(1,0)=\staubd$ and $x,y\in\Sigma^*$. Then
there exists $K'$ such that 
\be\label{lowerboundbd1}
\bk{\sigma(x)\sigma(y)}_{\bbbl^*}\geq K'{\exp\{-\staubd(y-x)\}\over
\|x-y\|^{3/2}}\,.
\ee
\item
Let $\stau(1,0)>\staubd$ and  $x,y\in\Sigma^*$. Then
there exists  $C>0$ such that
\be\label{lowerbound22}
\bk{\sigma(x)\sigma(y)}_{\bL^*}\geq C\exp\{-\staubd\|x-y\|\}\,.
\ee
\end{enumerate}
\end{pro}

\begin{pro}\label{propv}
Let $\beta^*<\beta_c$ and $0<h^*<\infty$.
\begin{enumerate}
\item
Let $x$, $y$ be two different points of $\bZ^{2*}$. Then
\be
\bk{\sigma(x)\sigma(y)}=\sum_{\lambda:x\ra y}
q^*(\lambda)\leq \exp\{-\stau(y-x)\}\,.
\ee
\item
Let $x$, $y$ be two different points of $\Sigma^*_L$. Then
\be
\bk{\sigma(x)\sigma(y)}_{\Lambda_L^*}\leq
\bk{\sigma(x)\sigma(y)}_{\bL^*}=
\sum_{\lambda:x\ra y}
q^*_{\bL^{*}}(\lambda)\leq \exp\{-\staubd\cdot\|y-x)\|\}\,.
\ee
\item
Let $x$, $y$ be two different points of $\Lambda_L^*$. Then
\be
\sum_{{\lambda:x\ra y\atop\lambda\cap\cE^*(\Sigma^*)=\emptyset}}
q_L^*(\lambda;\beta^*,h^*)\leq 
\bk{\sigma(x)\sigma(y)}(\beta^*)\,.
\ee
\item 
Let $x$, $y$, $z$ be three different points of $\Lambda_L^*$. Then
\be
\sum_{{\sc\lambda:x\ra z\atop\sc y\ni\lambda}}
q_L^*(\lambda)\leq 
\sum_{\lambda:x\ra y}
q_L^*(\lambda)
\sum_{\lambda:y\ra z}
q_L^*(\lambda)\,.
\ee
\end{enumerate}
\end{pro}

{\bf Remark:} Statement 4. of Proposition \ref{propv}
can be written as
\be
\sum_{{\sc\lambda:x\ra z\atop\sc y\ni\lambda}}
q_L^*(\lambda)\leq 
\bk{\sigma(x)\sigma(y)}_{\Lambda_L^*}
\,\bk{\sigma(y)\sigma(z)}_{\Lambda_L^*}\,.
\ee
There are variants of 4.. Let $\lambda:x\ra z$
and $y\in\lambda$; let $\lambda_1$ be the part of $\lambda$
from $x$ to $y$ and $\lambda_2$ be the part of $\lambda$
from $y$ to $z$. Then
\be
\sum_{{\sc\lambda:x\ra z\,, y\ni\lambda
\atop\sc \lambda_1\cap\cE^*(\Sigma^*)=\emptyset}}
q_L^*(\lambda;\beta^*,h^*)\leq 
\bk{\sigma(x)\sigma(y)}_{\Lambda_L^*}(\beta^*,1)
\,\bk{\sigma(y)\sigma(z)}_{\Lambda_L^*}(\beta^*,h^*)\,.
\ee
or
\be
\sum_{{{\sc\lambda:x\ra z\,, y\ni\lambda
\atop\sc \lambda_1\cap\cE^*(\Sigma^*)=\emptyset}\atop\sc
\lambda_2\cap\cE^*(\Sigma^*)=\emptyset}}
q_L^*(\lambda;\beta^*,h^*)\leq 
\bk{\sigma(x)\sigma(y)}_{\Lambda_L^*}(\beta^*,1)
\,\bk{\sigma(y)\sigma(z)}_{\Lambda_L^*}(\beta^*,1)\,.
\ee
\bigskip

We prove Propositions \ref{Ellipse_bulk} to
\ref{Ellipse_partial}, which state  concentration
properties of the  random  lines contributing to the two-point
function. Given $x$, $y\in\bZ^{2*}$ and $\rho>0$ we set
\be
\cS(x,y;\rho):=\{t \in \bZ^{2*}\,:\,\|t-x\|+\|t-y\|\leq \|y-x\| +
\rho\} \,,
\ee
and $\partial \cS(x,y;\rho)$ is the set of 
$z\in\bZ^{2*}\backslash\cS(x,y;\rho)$ such that there exists an edge 
$\bk{z,z'}$ with $z'\in\cS(x,y;\rho)$.

\begin{pro}\label{Ellipse_bulk}
Let $x$, $y\in\Lambda_L^*$ and $\rho>0$. Let $\cS_1:=\cS(x,y;\rho)$.
Then ($\Delta$ is defined in (\ref{striangle}) and (\ref{stiffness}))
\be\label{sss}
\sum_{{{\sc \lambda:\,x\mapsto y \atop\sc
\lambda\not\subset\cE^*(\cS_1)}
\atop\sc\lambda\cap\cE^*(\cS_1)\cap\cE^*(\Sigma^*_L)=\emptyset}}
q_L^*(\lambda) 
\leq 
|\partial \cS_1| \exp\{-\Delta\rho\}\exp\{-\stau(y-x)\}
\ee
$$
+\sum_{z_1,z_2\in\Sigma^*_L}
\exp\{-\stau(z_1-x)\}\exp\{-\staubd(z_2-z_1)\}\exp\{-\stau(y-z_2)\}\,.
$$
\end{pro}
\prf
Let $s\mapsto\lambda(s)$ be a parameterization of the open
contour $\lambda$ from $x$ to $y$. Let $s_1$ be the first time
(if any) such that $\lambda(s_1)\in\Sigma^*_L$ and $s_2$
the last time such that $\lambda(s_2)\in\Sigma^*_L$.
\bea
\sum_{{{\sc \lambda:\,x\mapsto y \atop\sc
\lambda\not\subset\cE^*(\cS_1)}
\atop\sc\lambda\cap\cE^*(\cS_1)\cap\cE^*(\Sigma^*_L)=\emptyset}}
q_L^*(\lambda) 
&\leq& 
\sum_{t\in \partial \cS_1} \sum_{{\sc\lambda:
x\rightarrow y\,, \lambda \ni t\atop\sc
\lambda \cap\cE^*(\Sigma^*_L)=\emptyset}}
q_L^*(\lambda)\\
&+&
\sum_{z_1,z_2\in \Sigma^*_L} \sum_{{\sc\lambda:
x\rightarrow y\atop\sc
\lambda(s_1)=z_1,\lambda(s_2)=z_2}} q_L^*(\lambda)\nnm\,.
\eea
Using Proposition \ref{propv}, the remark following it and 
the sharp triangle inequality, we get
\bea\label{avant}
\sum_{t\in \partial \cS_1} \sum_{{\sc\lambda:
x\rightarrow y\,, \lambda \ni t\atop\sc
\lambda \cap\cE^*(\Sigma^*_L)=\emptyset}}
q_L^*(\lambda)
&\leq& \sum_{t\in \partial \cS_1}
\bk{\sigma(x)\sigma(t)}_{\Lambda_L^*}(\beta^*,1)\,
\bk{\sigma(t)\sigma(y)}_{\Lambda_L^*}(\beta^*,1)\\
&\leq& 
\sum_{t\in \partial \cS_1}
\exp\{-(\stau(t-x)+\stau(y-t))\} \nnm\\
&\leq& \sum_{t\in \partial \cS_1} \exp\{-(\stau(y-x)+\Delta\rho)\}
\nnm\\
&\leq& |\partial \cS_1| \exp\{-\Delta\rho\}
\exp\{-\stau(y-x)\}\,,\nnm
\eea
and
\bea
\sum_{z_1,z_2\in \Sigma^*_L} \sum_{{\sc\lambda:
x\rightarrow y\atop\sc
\lambda(s_1)=z_1,\lambda(s_2)=z_2}} q_L^*(\lambda)
&\leq&
\sum_{z_1,z_2\in\Sigma^*_L}
\exp\{-\stau(z_1-x)\}\\
& &\cdot
\exp\{-\staubd(z_2-z_1)\}\exp\{-\stau(y-z_2)\}\,.\nnm
\eea
\qed

{\bf Remarks.}

1. If $h^*\leq 1$, then the statement (\ref{sss}) simplifies,
\be\label{4.17}
\sum_{ {\sc\lambda:x\mapsto y\atop\sc 
\lambda\not\subset\cE^*(\cS_1)}}
q_L^*(\lambda) 
\leq 
|\partial \cS_1| \exp\{-\Delta\rho\}\exp\{-\stau(y-x)\}\,.
\ee

2. At the thermodynamical limit we can improve (\ref{4.17})
using Proposition \ref{lowerboundpro}. 
\be\label{4.18}
\sum_{ {\sc\lambda:x\mapsto y\atop\sc 
\lambda\not\subset\cE^*(\cS_1)}}
q^*(\lambda) 
\leq 
{1\over K}|\partial \cS_1| \|x-y\|^{1/2}
\exp\{-\Delta\rho\}\bk{\sigma(x)\sigma(y)}\,.
\ee
The replacement of $\exp\{-\stau(y-x)\}$ by $\bk{\sigma(x)\sigma(y)}$
is significant since $\bk{\sigma(x)\sigma(y)}$ is the total mass
of the measure defined by $q^*(\lambda) $ on the set of all
$\lambda$ with $\delta\lambda=\{x,y\}$.

3. If $1<h^*<h^*_c$, then the sharp triangle inequality applied
two times to (\ref{sss}) gives
\be
\sum_{{{\sc \lambda:\,x\mapsto y \atop\sc
\lambda\not\subset\cE^*(\cS_1)}
\atop\sc\lambda\cap\cE^*(\cS_1)\cap\cE^*(\Sigma^*_L)=\emptyset}}
q_L^*(\lambda) 
\leq 
\exp\{-\stau(y-x)\}\Big(|\partial \cS_1| \exp\{-\Delta\rho\}
\ee
$$
+\sum_{z_1,z_2\in\Sigma^*_L}
\exp\{-\Delta(\|z_1-x\|+\|z_2-z_1\|+\|y-z_2\|-\|x-y\|)\}\Big)\,.
$$

\begin{pro}\label{Ellipse_complete}
Let $x$, $y\in\Lambda_L^*$ with $x(2)=y(2)=-1/2$, and $\rho>0$. Let 
$\cS_2 :=\cS(x,y;\rho)$.  If $\stau(1,0)=\staubd$, that is
$h^*\leq h^*_c$, then 
\be
\sum_{{\sc\lambda :x\mapsto y\atop\sc
\lambda\not\subset\cE^*(\cS_2)}}q^*_L(\lambda) \leq
O(\|x-y\|+\rho)
|\partial \cS_2| \exp\{-\Delta\rho\}\exp\{-\stau(y-x)\}\,.
\ee
\end{pro}
\prf
If $\stau(1,0)=\staubd$ and $u$, $v\in\bbbl^*$ with $u(2)=v(2)=-1/2$,
then
\be\label{equality}
\staubd \|u-v\|=\stau(u-v)\,.
\ee
Let $\lambda:x\mapsto y$, with $\lambda\ni t$, $t\in\partial \cS_2$.
We consider $\lambda$ as a parameterized curve, 
$s\mapsto\lambda(s)$, from $x$ to $y$. We set $t=\lambda(s^*)$; we
denote by $s_1$ the last time before $s^*$ such that 
$\lambda(s_1)\in\Sigma^*_L$;  we denote by $s_2$ the
first time after $s^*$ such that  $\lambda(s_2)\in\Sigma^*_L$.
We have
\bea
\sum_{\lambda:x\rightarrow y \atop \lambda \not\subset\cE^*(\cS_2)} 
q^*_L(\lambda) &\leq& \sum_{t\in \partial \cS_2} \sum_{\lambda:
x\rightarrow y \atop \lambda \ni t}
q^*_L(\lambda)\\
&\leq& \sum_{t\in \partial \cS_2} \sum_{u,v\in\Sigma^*_L}
\sum_{{\sc\lambda: x\rightarrow y\,,t\in\lambda \atop \sc
\lambda(s_1)=u,\lambda(s_2)=v}}q^*_L(\lambda)\,.\nnm
\eea
Using (\ref{equality}), Proposition \ref{propv},
GKS inequalities and  the sharp
triangle inequality, we get 
\bea
\sum_{u,v}\sum_{\lambda:
x\rightarrow y \atop \lambda \ni u,t,v}
q^*_L
&\leq&\sum_{u,v}
\bk{\sigma(x)\sigma(u)}_{\bbbl^*}\bk{\sigma(u)\sigma(t)}
\bk{\sigma(t)\sigma(v)}\bk{\sigma(v)\sigma(y)}_{\bbbl^*}
\nnm\\
&\leq& \sum_{u,v}
\exp\{-\stau(u-x)-\stau(t-u)\}
\exp\{-\stau(v-t)-\stau(y-v)\}\nnm\\ 
&\leq&\sum_{u,v}
\exp\{-\stau(t-x)\}\exp\{-\Delta(\|x-u\|+\|u-t\|-\|x-t\|)\}\nnm\\
& &
\cdot\exp\{-\stau(y-t)\}\exp\{-\Delta(\|y-v\|+\|v-t\|-\|y-t\|)\}\nnm\\
&\leq&
O(\|x-y\|+\rho)\exp\{-(\stau(t-x)+\stau(y-t))\}\,.\nnm
\eea
Then the proof is as in (\ref{avant}).
\qed

In the case of partial wetting, 
i.e. $\stau(1,0)>\staubd(1,0)$, the previous proposition can be
improved to reflect the fact that the contours stick
to the wall, even microscopically. 

\begin{pro}\label{Ellipse_partial}
Suppose $h^*>h_c^*$ (partial wetting for the dual model). Let $x$,
$y\in\Lambda^*_L$ with $x(2)=y(2)=-1/2$,
$x(1)<y(1)$, and let $\rho_i>0$, $i=1,2$. Let
\be
\cS_3:= \{t \in \Lambda^*_L\,:\,x(1)-\rho_1\leq t(1) \leq y(1)+
\rho_1\,,\, -1/2\leq t(2)\leq \rho_2\}\,.
\ee
Then, there exists a constant $\overline{C}(\beta)>0$ such that
\be
\sum_{{\sc\lambda :x\rightarrow y\atop\sc
\lambda \not\subset\cE^*(\cS_3)}}q^*_L(\lambda) \leq
|\partial \cS_3|
\Big(\exp\{-2\rho_1\staubd\}+|\rho_2|\exp\{-\overline{C}\rho_2\}\Big)
\exp\{-\staubd\|y-x\|\}\,. 
\ee
\end{pro}
\prf
Let
$\partial \cS_3 := \{t\in\cS_3 \,:\, t(2)=\rho_2\mbox{ or } 
t(1) =x(1)-\rho_1 \mbox{ or } t(1) = y(1)+\rho_1\}$; 
we can write
\bea
\sum_{\sc\lambda:x\rightarrow y 
\atop \sc \lambda \not\subset\cE^*(\cS_3)}
q^*_L(\lambda)
&\leq& \sum_{t\in \partial \cS_3} 
\sum_{\sc\lambda: x\rightarrow y 
\atop \sc\lambda \ni t}
q^*_L(\lambda)\\
&\leq& \sum_{\sc t\in \partial \cS_3 \atop\sc t(2)<\rho_2}
\sum_{\sc\lambda: x\rightarrow y \atop \sc\lambda \ni t} 
q^*_L(\lambda)
+ \sum_{\sc t\in \partial \cS_3 \atop\sc  t(2)=\rho_2}
\sum_{\sc\lambda: x\rightarrow y \atop \sc\lambda \ni t} 
q^*_L(\lambda)\,.\nnm
\eea
We treat these  sums separately. By symmetry and GKS inequalities
\bea
\sum_{t\in \partial \cS_3 \atop t(2)<\rho_2}
\sum_{\lambda: x\rightarrow y \atop
\lambda \ni t} q^*_L(\lambda) 
&\leq& 
2\sum_{\sc t\in \partial \cS_3 \atop\sc
t(1)=x(1)-\rho_1}\bk{\sigma_x\sigma_t}_{\bbbl^*}
\bk{\sigma_t\sigma_y}_{\bbbl^*}\\
&=& 2\sum_{\sc t\in \partial \cS_3 \atop\sc t(1)=x(1)-\rho_1}
\bk{\sigma_{\overline{x}}\sigma_t}_{\bbbl^*}
\bk{\sigma_t\sigma_y}_{\bbbl^*}\nnm\\
&\leq&2 \sum_{\sc t\in \partial \cS_3 \atop\sc t(1)=x(1)-\rho_1}
\bk{\sigma_{\overline{x}}\sigma_y}_{\bbbl^*}\nnm\\
&\leq& 2\sum_{\sc t\in \partial \cS_3 \atop\sc t(1)=x(1)-\rho_1}
\exp\{-2\rho_1\staubd\}\exp\{-\staubd\|y-x\|\}\,,\nnm
\eea
where $\overline{x}$ is the image of $x$ under a reflection of axis
$\{u\,:\,u(1)=t(1)\}$. 

Let $t\in\lambda$ and $t(2)=\rho_2$. As above
$\lambda$ is considered as a parameterized curve, and we set
$t=\lambda(s^*)$; we denote by $s_1$ the last time before $s^*$ such
that  $\lambda(s_1)\in\Sigma^*_L$;  we denote by $s_2$
the first time after $s^*$ such that  $\lambda(s_2)\in\Sigma^*_L$.
As above we get ($u=\lambda(s_1)$, $v=\lambda(s_2)$)
\bea
\sum_{\sc t\in \partial \cS_3 \atop \sc t(2)=\rho_2}
\sum_{\lambda: x\rightarrow y \atop
\lambda \ni t} q^*_L(\lambda) 
&\leq& 
\sum_{\sc t\in \partial \cS_3 \atop\sc t(2)=\rho_2}
\sum_{u,v}
\bk{\sigma_x\sigma_u}_{\bbbl^*}  \bk{\sigma_u\sigma_t}
\bk{\sigma_t\sigma_v} \bk{\sigma_v\sigma_y}_{\bbbl^*} \\
&\leq& \sum_{\sc t\in \partial \cS_3 \atop \sc t(2)=\rho_2}
\sum_{u,v} \exp\{-\staubd(\|u-x\|+\|y-v\|)\}\nnm\\
& &\cdot\exp\{-\stau(t-u) - \stau(v-t)\}\nnm\\ 
&\leq& 
\sum_{\sc t\in \partial \cS_3 \atop \sc t(2)=\rho_2}
\sum_{u,v} \exp\{-\staubd(\|u-x\|+\|y-v\|)\}\nnm\\
& &
\cdot\exp\{-\stau(u-v)\}
\exp\{-\Delta(\|u-t\|+\|t-v\|-\|u-v\|)\}\nnm\\
&\leq& 
\sum_{\sc t\in \partial \cS_3 \atop \sc t(2)=\rho_2}
\sum_{u,v} \exp\{-\staubd\|x-y\|\}
\exp\{-(\stau(1,0)-\staubd)\|u-v\|\}\nnm\\
& &\cdot\exp\{-\Delta(\|u-t\|+\|t-v\|-\|u-v\|)\}\nnm\,,
\eea
The conclusion follows from the observations: 1. the summation is over the
base of the triangle $uvt$; 2.  the term
$\exp\{-(\stau(1,0)-\staubd)\|u-v\|\}$ allows to control the triangles with a large base, while the term
$\exp\{-\Delta(\|u-t\|+\|t-v\|-\|u-v\|)\}$ can be used to control the terms in
which the base is far from the point $t$.
\qed


\section{Probability of the phase separation line}\label{Estimate}
\setcounter{equation}{0}

We study the probability of the phase separation line by making
a coarse--grained description of it.
We estimate in terms of its surface
tension\footnote{
In our problem it is
sufficient to give an extremely rough description of the
phase separation line, because we do not need to control the volume
under the phase separation line, as it was the case in \cite{PV1}.}
the probability that a
given coarse--grained description occurs.

We first prove an essential lower bound and then proceed with the 
main estimate.
\begin{pro}\label{lowerbd}
Let $\beta>\beta_c$, $h>0$, $0<a<1$, $0<b<1$ and $\W^*=\W^*(\beta,h)$
be the minimum  of the functional $\W$.  Then there exists $C>0$
and $L_0=L_0(\beta,h)$ such
that, for all $L\geq L_0$,
\be
Z^{ab}(\Lambda_L) \geq {\exp\{-\W^*L\}\over L^C}
Z^{-}(\Lambda_L)\,. 
\ee
\end{pro}

{\bf Remark.} The dual statement of Proposition \ref{lowerbd} is
\be
\bk{\sigma(t_l^L)\sigma(t_r^L)}_{\Lambda_L^*}\geq
{\exp\{-\W^*L\}\over L^C}\,.
\ee

\prf
We can write, using Proposition \ref{prorandom},
\be
Z^{ab}(\Lambda_L) = Z^{-}(\Lambda_L)
\sum_{\sc \lambda: \atop \sc\delta\lambda =
\{t_l^L,t_r^L\}}q^*_{L}(\lambda)\,.
\ee
Let $\cC^*$ be the simple rectifiable curve in $Q$ which realizes 
the minimum of the variational problem (or one of the minima in
case of degeneracy). Let $K_1>0$, which will be chosen large enough
below.

We first consider the case $\cC^* = \cD$.  Let $u^L_l$ and $u^L_r$ 
be the points of $\Lambda_L^*$ with 
$u_l^L(1)=t_l^L(1)+[K_1\log L]$, $u_r^L(1)=t_r^L(1)-[K_1\log L]$,
which are closest to the straight line from $t_l^L$ to
$t_r^L$.

We need the following result 

\begin{lem}\label{lem5.1}
Let $\cB$ be a rectangular box in $\Lambda_L^*$, and $x$, $y$ two points on its
boundary. Let $d>0$ and $u$, $v$ be two points in $\cB$ such that $u$ and $v$ are closest
to the straight line from $x$ to $y$ and $\|x-u\|=\|y-v\|=d$.
If the distance of $\cB$ to $\Sigma^*_L$ is larger than $C'\log L$,
with $C'>2$, then
\be
\sum_{\lambda:x\ra y\atop\lambda\subset \cE^*(\cB)}
q^*_L(\lambda)\geq
\exp\{-O(\log L)\}
\sum_{\lambda:u\ra v\atop\lambda\subset \cE^*(\cB)}
q^*(\lambda)\,.
\ee
\end{lem}

\prf
Following the proof of Proposition 6.1 of \cite{PV1}, 
from (6.38) to (6.42), we get
\be
\sum_{\lambda:x\ra y\atop\lambda\subset \cE^*(\cB)}
q^*_L(\lambda)\geq
\exp\{-O(\log L)\}
\sum_{\lambda:u\ra v\atop\lambda\subset \cE^*(\cB)}
q^*_L(\lambda)\,.
\ee
Since the distance of $\cB$ to $\Sigma^*$ is larger than $C'\log L$
and  $C'>2$, then it follows from point 4. of Lemma 5.3  in
\cite{PV1} that there exists a constant $\tilde{C}>0$
such that for $L$ large enough and all 
$\lambda \subset \cE^*(\cB)$
\be\label{5.123}
q^*_L(\lambda)\geq q^*_{\bL^*}(\lambda)\geq \tilde{C}q^*(\lambda)\,.
\ee
This proves the lemma. \qed

\begin{figure}
 \centerline{\psfig{figure=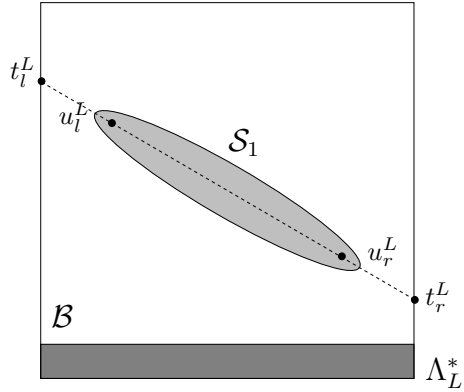,height=50mm}}
  \figtext{
 	\writefig	7.35	3.55	{$\cS_1$}
 	\writefig	9.97	0.50	{$\Lambda^*_L$}
 	\writefig	5.0	1.2	{$\cB$}
 	\writefig	4.45	4.50	{\footnotesize $t_l^L$}
 	\writefig	9.97	1.5	{\footnotesize $t_r^L$}
 	\writefig	9.2	2.1	{\footnotesize $u_r^L$}
 	\writefig	5.1	3.9	{\footnotesize $u_l^L$}
  }
 \caption[]{The ellipse $\cS_1$; the box $\cB$ is the whole box minus the
 bottom strip.}
 \label{fig_S1}
\end{figure}

Let $C'>2$.
We introduce $\cB := \{t\in\Lambda_L^*\,:\,t(2)>[C'\log L]\}$.
Let $\cS_1$ be the elliptical
set of Proposition \ref{Ellipse_bulk} with $x=u_l^L$, $y=u_r^L$,
$\rho=O(K_1\log L)$ so that 
$\cE^*(\cS_1)\subset\cE^*(\cB)$ (see Fig. \ref{fig_S1}). From Lemma \ref{lem5.1}, applied to the box
$\cB$  with $x=t_l^L$, $y=t_r^L$ , $u=u_l^L$ and $v=v_l^L$,
and the second remark following Proposition
\ref{Ellipse_bulk}, we get 
\bea
{Z^{ab}(\Lambda_L)\over Z^{-}(\Lambda_L)} 
&\geq&
\sum_{\lambda:t_l^L\ra t_r^L\atop\lambda\subset \cE^*(\cB)}
q^*_L(\lambda)\\
&\geq&
\exp\{-O(\log L)\}
\sum_{\lambda:u_l^L\ra u_r^L\atop\lambda\subset \cE^*(\cS_1)}
q^*(\lambda)\nnm\\
&\geq&
\exp\{-O(\log L)\}\Big(\sum_{\lambda:u_l^L\ra u_r^L}q^*(\lambda)
-\sum_{\lambda:u_l^L\ra u_r^L\atop\lambda\not\subset\cE^*(\cS_1)}
q^*(\lambda)\Big)\nnm\\
&\geq&
\exp\{-O(\log L)\}\bk{\sigma(u_l^L)\sigma(u_r^L)}\Big(1-
{1\over K}|\partial\cS_1|\,\|u_l^L-u_r^L\|^{1/2}\exp\{-\Delta\rho\}
\Big)
\nnm\\
&\geq& 
{\exp\{-\stau(t_l^L-t_r^L)\}\over L^C}\,,\nnm
\eea
for some positive constant $C$, by taking $K_1$ large enough.

We now consider the case $\cC^*=\cW$. Since $h>0$, the angle 
$\theta_Y$ satisfies $0<\theta_Y<\pi/2$.  Denote by $w_1^L$ and
$w_2^L$  the two points on $\Sigma^*_L$ 
which are closest to the corners of the  polygonal line
$\cW$ scaled by $L$.

\begin{figure}[thb]
 \centerline{\psfig{figure=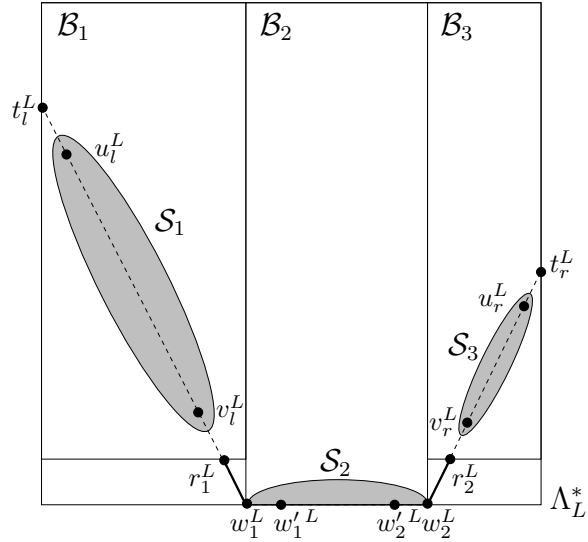,height=70mm}}
  \figtext{
 	\writefig	7.38	0.50	{\footnotesize ${w'_1}{}^L$}
 	\writefig	6.62	2.02	{\footnotesize $v_l^L$}
 	\writefig	8.75	0.50	{\footnotesize ${w'_2}{}^L$}
 	\writefig	9.45	1.85	{\footnotesize $v_r^L$}
 	\writefig	3.9	6.00	{\footnotesize $t_l^L$}
 	\writefig	11.05	3.95	{\footnotesize $t_r^L$}
 	\writefig	9.35	0.50	{\footnotesize $w_2^L$}
 	\writefig	5.0	5.50	{\footnotesize $u_l^L$}
 	\writefig	10.1	3.50	{\footnotesize $u_r^L$}
 	\writefig	6.8	0.50	{\footnotesize $w_1^L$}
 	\writefig	11.05	0.8	{$\Lambda_L^*$}
 	\writefig	4.5	7.10	{$\cB_1$}
 	\writefig	7.2	7.10	{$\cB_2$}
 	\writefig	9.6	7.10	{$\cB_3$}
 	\writefig	5.8	4.50	{$\cS_1$}
 	\writefig	8.0	1.30	{$\cS_2$}
 	\writefig	9.7	2.85	{$\cS_3$}
 	\writefig	9.75	1.10	{\footnotesize $r_2^L$}
 	\writefig	6.25	1.10	{\footnotesize $r_1^L$}
  }
 \caption[]{The three boxes $\cB_i$, $i=1,\dots,3$ and their elliptical
 subsets; the two bold segments represent the two shortest contours $\overline{\lambda}_i$,
 $i=1,2$.
 }
 \label{fig_S1S2S3}
\end{figure}

We define three rectangular boxes (see Fig. \ref{fig_S1S2S3})
\bea
{\cB}_1 &=& \{t\in \Lambda_L^*\,:\,t(1)\leq w_1^L(1)\,,\,t(2)>[C'\log L]\}\,,\\
{\cB}_2 &=& \{t\in \Lambda_L^*\,:\,w_1^L(1)\leq t(1)\leq w_2^L(1)\}\,,\\
{\cB}_3 &=& \{t\in \Lambda_L^*\,:\,w_2^L(1)\leq t(1)\,,\,t(2)>[C'\log L]\}\,.
\eea
Moreover let $r_1^L$, resp. $r_2^L$, be the point of $\Lambda_L^*$ closest to
the straight line through $t_l^L$ and $w_1^L$, resp. $t_r^L$ and $w_2^L$, 
such that $r_1^L(2) = [C'\log L]+1/2$, resp. $r_2^L(2) = [C'\log L]+1/2$.
Let $\overline{\lambda}_1$, resp. $\overline{\lambda}_2$, be a shortest open
contour from $r^L_1$ to $w^L_1$, resp. from $r^L_2$ to $w^L_2$.
We define $\cA$ as the set $\cA$ of open contours
$\lambda=
\lambda_1\cup\overline{\lambda}_1\cup\lambda_2\cup\overline{\lambda_2}
\cup\lambda_3$
such that
\begin{itemize}
\item $\lambda_i \subset \cE^*(\cB_i)$, $i=1,\dots,3$;
\item $\lambda_1$ : $t_l^L \ra r_1^L$;
\item $\lambda_2$ : $w_1^L \ra w_2^L$;
\item $\lambda_3$ : $r_2^L \ra t_r^L$.
\end{itemize}
We can write
\bea
{Z^{ab}(\Lambda_L)\over Z^{-}(\Lambda_L)} &\geq& 
\sum_{\lambda\in \cA} q^*_{L}(\lambda)\nnm\\
&\geq& \exp\{-O(\log L)\}\prod_i
\sum_{\sc\lambda_i\subset \cE^*(\cB_i): 
\atop\sc
\lambda_i\,\mbox{\scriptsize as above}} q^*_{L}(\lambda_i)\,.
\eea
We apply Lemma \ref{lem5.1} to the sum over $\lambda_1$ with $\cB=\cB_1$, $x=t_l^L$,
$y=r_1^L$, $d=K_1\log
L$; we denote by $u_l^L$ and $v_l^L$ the corresponding points $u$ and $v$.
We apply the same lemma to the sum over $\lambda_3$ with $\cB=\cB_2$, $x=r_2^L$, $y=t_r^L$, $d=K_1\log
L$; we denote by $u_r^L$ and $v_r^L$ the corresponding points $u$ and $v$.
The sum over $\lambda_2$ is taken care of by the following

\begin{lem}\label{lem5.2}
Let ${w_1'}{}^L := w_1^L+([K_1\log L], 0)$, $w'_2{}^L := w_2^L-([K_1\log L], 0)$. Then
\be
\sum_{\lambda:w_1^L\ra w_2^L\atop\lambda\subset \cE^*(\cB_2)}
q^*_L(\lambda)\geq
\exp\{-O(\log L)\}
\sum_{\lambda:{w'_1}{}^L\ra {w'_2}{}^L\atop\lambda\subset \cE^*(\cB_2)}
q_{\bL^*}^*(\lambda)\,.
\ee
\end{lem}

\prf
Same proof of that of Proposition 6.2 in \cite{PV1}. \qed 

We finally introduce three elliptical sets. $\cS_1$ is constructed as in
Proposition \ref{Ellipse_bulk} with $x=u_l^L$, $y=v_l^L$ and $\rho=O(K_1\log L)$ such that
$\cE^*(\cS_1) \subset \cE^*(\cB_1)$; $\cS_3$ is constructed as in
Proposition \ref{Ellipse_bulk} with $x=v_r^L$, $y=u_r^L$ and $\rho=O(K_1\log L)$ such that
$\cE^*(\cS_3) \subset \cE^*(\cB_3)$; $\cS_2$ is constructed as in
Proposition \ref{Ellipse_complete} with $x={w'_1}{}^L$, $y={w'_2}{}^L$ and $\rho=[K_1\log L]$ (and
therefore $\cE^*(\cS_2) \subset \cE^*(\cB_2)$).

Applying Propositions \ref{Ellipse_bulk} and \ref{Ellipse_complete} as before gives the conclusion,
\be 
{Z^{ab}(\Lambda_L)\over Z^{-}(\Lambda_L)} \geq {\exp\{-L\cW^*\}\over L^C}\,
\ee
for some positive constant $C$.\qed

We prove using the above proposition, that the 
surface tension of a (very) coarse-grained version
of the  phase separation line cannot be too large 
compared to $\W^*$.

Let $\lambda$ be the open contour. We construct a polygonal line 
approximation $\cP :=\cP(\lambda)$ of $\lambda$. Let $s\mapsto
\lambda(s)$ be a unit-speed parameterization of $\lambda$. If
$\lambda(s)(2)>-1/2$ for all $s$, then let 
$\cP$ be the straight line
from $t_l^L$ to $t_r^L$. Otherwise, let $s_1$  
be the first time such
that $\lambda(s)(2)=-1/2$ and $s_2$ the  last  time such that
$\lambda(s)(2)=-1/2$;  we write 
$\hat{w}_i^L = \lambda(s_i)$, $i=1,2$.
We also introduce $[\cP] := \{\omega\,:\,\cP(\lambda(\omega)) =
\cP$\}.

By construction, if $s<s_1$ or $s>s_2$ then 
$\lambda(s)(2)\neq -1/2$. We can therefore apply Proposition 
\ref{propv}
to estimate the probability of the event $[\cP]$,
\be\label{5.6}
P_L^{ab}[\,[\cP]\,] \leq \exp\{-\W(\cP)\}\,.
\ee

\begin{pro}\label{decroissance}
Let $\beta>\beta_c$, $h>0$, $0<a<1$, $0<b<1$. Then there exists
$L_0=L_0(\beta,h)$ such that, for all $L\geq L_0$ and $T>0$,
\be
P_L^{ab}[\,\{\W(\cP(\lambda))\geq \W^*L+T\}\,] \leq 
\exp\{-T + O(\log L)\}\,.
\ee
\end{pro}
\prf
Let 
\be
\cI(T) := 
\{\lambda\subset \cE^*(\Lambda_L^*) \,:\, \delta\lambda =
\{t_l^L,t_r^L\}\,,\,\W(\cP(\lambda))\geq \W^*L+T\}\,.
\ee
Then, from Propositions \ref{lowerbd}, \ref{prorandom} and
(\ref{5.6}),
\bea
P_L^{\pm}[\,\{\W(\cP(\lambda))\geq \W^*L+T\}\,] &=&
{Z^{-}(\Lambda_L)\over Z^{ab}(\Lambda_L)}\sum_{\lambda\in\cI(T)} 
q^*_L(\lambda)\\
&\leq& \exp\{\W^*L+O(\log L)\} \sum_{\lambda\in\cI(T)}
q^*_L(\lambda)
\nnm\\
&\leq& \exp\{\W^*L+O(\log L)\} \sum_{\sc\cP: \atop\sc
\W(\cP)\geq\W^*L+T} \sum_{\sc\lambda: \atop \sc\cP(\lambda)=\cP}
q^*_L(\lambda)\nnm\\
&\leq& \exp\{\W^*L+O(\log L)\} \sum_{\sc\cP: \atop\sc
\W(\cP)\geq\W^*L+T} \exp\{-\W(\cP)\}\nnm\\
&\leq& \exp\{-T+O(\log L)\}\,.\nnm
\eea
(The number of different coarse--grained polygonal lines is
bounded by $O(L^2)$.)
\qed


\section{Pinning transition}\label{Continuum}
\setcounter{equation}{0}

The main result of the paper is a statement about concentration
properties of the probability of the phase separation line. An
immediate consequence of Theorem \ref{mainthm} is that at a
suitable scale, when $L\ra \infty$, the phase separation line 
defines a non--random object, the interface, which is characterized
as the solution of the variational problem discussed in section 
\ref{Variational}, that is, the interface in $Q$ is either
the straight line $\cD$ or the broken line $\cW$.
We obtain an
essentially optimal description of the location of the 
interface up to
the scale of normal fluctuations of the phase separation line.

\subsection{Main result}\label{mainresult}

The weight of a separation line $\lambda$ in $\Lambda_L^*$,
going from $t_l^L$ to $t_r^L$, is given by
$q_L(\lambda;\beta,h)=q_L^*(\lambda;\beta^*,h^*)$. These weights
define a measure on the the set of the phase
separation lines, such that the total mass is
\be
\sum_{\sc \lambda\subset\cE(\Lambda_L^*):\atop
\sc \delta\lambda=\{t_l^L,t_r^L\}} q_L^*(\lambda;\beta^*,h^*)=
\bk{\sigma(t_l^L)\sigma(t_r^L)}_{\Lambda_L^*}(\beta^*,h^*)\,.
\ee
Consequently
\be\label{proba}
P^{ab}_L[\lambda]={q_L^*(\lambda)\over
\bk{\sigma(t_l^L)\sigma(t_r^L)}_{\Lambda_L^*}}\,.
\ee

Let $\cD$ and $\cW$ be the curves in $Q$ introduced in section 
\ref{Variational}. We set 
\be\label{definterval}
I^L_i:=\{\,x\in\Sigma^*_L:\,\|x-w_i^L\|\leq 
(ML\log L)^{1/2}\,\}\quad ,\quad i=1,2\,,
\ee
and
\be\label{defrho}
\rho_L:=M\log L\,.
\ee 
We define two sets of contours.
The set ${\cT}_\cD$ contains all $\lambda\subset\cE^*(\Lambda_L^*)$
such that 
\begin{enumerate}
\item[$a_1.$] 
$\delta\lambda=\{t_l^L,t_r^L\}$;
\item[$a_2.$] 
$\lambda$ is inside $\cS(t_l^L,t_r^L;\rho_L)$.
\end{enumerate}
The set ${\cT}_\cW$ contains all
$\lambda\subset\cE^*(\Lambda_L^*)$, considered as parameterized
curves $s\mapsto \lambda(s)$, such that  
\begin{enumerate} 
\item[$b_1.$] 
$\delta\lambda=\{t_l^L,t_r^L\}$, $\lambda(0):=t_l^L$;
\item[$b_2.$] 
$\exists s_1$ such that $\lambda(s_1)\in I^L_1$ and
for all $s<s_1$, $\lambda(s)\cap\Sigma^*_L=\emptyset$;
\item[$b_3.$]
$\lambda_1:=\{\lambda(s):\, s\leq s_1\}$ is inside 
$\cS(t_l^L,\lambda(s_1);\rho_L)$;
\item[$b_4.$]
$\exists s_2$ such that $\lambda(s_2)\in I^L_2$ and
for all $s_2< s$, $\lambda(s)\cap\Sigma^*_L=\emptyset$;
\item[$b_5.$]
$\lambda_3:=\{\lambda(s):\, s_2\leq s\}$ is inside 
$\cS(\lambda(s_2),t_r^L;\rho_L)$;
\item[$b_6.$]
$\lambda_2:=\{\lambda(s):\,s_1\leq s_2\leq s\}$ is inside
$$
\{x\in\Lambda_L^*:\,x(2)\leq \rho_L\;,\;
\lambda(s)(1)-\rho_L\leq x(1)\leq
\lambda(s)(2) +\rho_L\}\,.
$$
\end{enumerate}

\begin{thm}\label{mainthm}
Let $\beta>\beta_c$, $h>0$, $0<a<1$, $0<b<1$. There exist $M>0$ and
$L_0=L_0(h,\beta,M)$ such that, for all $L\geq L_0$, the following 
statements are true.
\begin{enumerate}
\item
Suppose that the solution of the variational problem in $Q$ is the
curve $\cD$. Then
\be
P^{ab}_L[{\cT}_\cD]\geq
1-L^{-O(M)}\,.
\ee
\item
Suppose that the solution of the variational problem in $Q$ is the
curve $\cW$. Then
\be
P^{ab}_L[{\cT}_\cW]\geq
1-L^{-O(M)}\,.
\ee
\item
Suppose that the solution of the variational problem in $Q$ is the
either the curve $\cD$ or the curve $\cW$. Then
\be
P^{ab}_L[{\cT}_\cD\cup{\cT}_\cW]\geq 1-L^{-O(M)}\,.
\ee
\end{enumerate}
\end{thm}
{\bf Comment:} The results of Theorem \ref{mainthm}  are, in some
sense, optimal.  Indeed, at a finer scale we do not expect the
phase separation line to  converge to some non--random set, 
but rather
to some random process. It is  known that fluctuations 
of a phase separation line of length $O(L)$, which 
is not in contact
with  the wall, are $O(L^{1/2})$ (see \cite{Hi} and \cite{DH}).
On the other hand, if
the phase separation line is attracted by the wall on a length 
$O(L)$, then we expect that its excursions away from the 
wall have a
size typically bounded by $O(\log L)$.

\prf

{\bf 1.} Suppose that the minimum of the variational 
problem is given by $\cD$,
$\W(\cD)=\W^*$.
Let $\W^{**}$ be the minimum of the functional over 
all simple curves in $Q$,
with end--points $A$ and $B$, and which touch the wall $w_Q$. 
By hypothesis
there exists $\delta>0$ with $\W^{**}=\W^*+\delta$.

We set $\cS_1:=\cS(t_l^L,t_r^L;\rho_L)$; for $L$ large enough 
$\cS_1\cap\Sigma^*_L=\emptyset$ since $a>0$ and $b>0$.

We apply Proposition \ref{lowerbd}. We have
\bea
P^{ab}_L[\{\lambda\not\in{\cT}_\cD\}]&=&{1\over
\bk{\sigma(t_l^L)\sigma(t_r^L)}_{\Lambda_L^*}}
\sum_{\lambda\not\in{\cT}_\cD}q_L^*(\lambda)\\
&\leq & L^C\exp\{\W^*L\}\sum_{\lambda\not\in{\cT}_\cD}
q_L^*(\lambda)\nnm\,.
\eea
\nnm
We apply Proposition \ref{Ellipse_bulk}.
\be
\sum_{\lambda\not\in{\cT}_\cD}q_L^*(\lambda)
\leq
|\partial \cS_1| \exp\{-\Delta\rho\}\exp\{-\stau(t_r^L-t_l^L)\}
\ee
$$
+\sum_{z_1,z_2\in\Sigma^*_L}
\exp\{-\stau(z_1-t_l^L)\}\exp\{-\staubd(z_2-z_1)\}
\exp\{-\stau(t_r^L-z_2)\}\,.
$$
We can bound above the last sum by $O(L^2)\exp\{-L\W^{**}\}$. 
Therefore
\be
P^{ab}_L[\{\lambda\not\in{\cT}_\cD\}]\leq 
{O(L^{C+1})\over L^{\Delta M}}
+O(L^{C+2})\exp\{-L\delta\}\,.
\ee
This proves the first statement.

{\bf 2.} Suppose that the minimum of the variational 
problem is given by $\cW$,
$\W(\cW)=\W^*$. Then there exists $\delta>0$ such
that $\W(\cD)=\W^*+\delta$.
We estimate  $P^{ab}_L[\{\lambda\not\in{\cT}_\cW\}]$ in 
several steps.
Notice that condition $b_1$ is always satisfied.\\
$1.$ The  probability that condition $b_2$ is satisfied, but not
$b_3$, can be estimated by Proposition
\ref{Ellipse_bulk}; it is smaller than $O(L^{C+1})/L^{\Delta M}$.\\
$2.$ The  probability that condition $b_4$ is satisfied, but not
$b_5$, is estimated in the same way; it is smaller than 
$O(L^{C+1})/L^{\Delta M}$.\\ 
$3.$  The  probability that conditions $b_2$ and $b_4$ 
are satisfied,
but not $b_6$, can be estimated by
Proposition \ref{Ellipse_partial}; it is smaller than 
$O(L^{C+1})/L^{\overline{C}\Delta M}$.\\ 
$4.$ We estimate the probability that condition $b_2$ is not 
satisfied.
The case with condition $b_5$ is similar. If 
$\lambda$ does not intersect
$\Sigma^*_L$, then this probability is smaller than 
$O(L^{C})\exp\{-\delta L\}$,
since $\W(\cD)=\W^*+\delta$.
Suppose  that there exist $s_1$ and $s_2$, with
$\lambda(s_i)\in\Sigma^*_L$, $\lambda(s)\cap\Sigma^*_L=\emptyset$ 
for all $s<s_1$
and $\lambda(s)\cap\Sigma^*_L=\emptyset$ for all $s_2<s$.  Let
$p^L_i:=\lambda(s_i)$, $i=1,2$. Under these conditions, $b_2$ 
is not satisfied
if and only if $p^L_1\not\in I^L_1$. Let $\cC(p^L_1,p^L_2)$ 
be the polygonal
curve from $t_l^L$ to $p^L_1$, then from $p^L_1$ to $p^L_2$ 
and finally from
$p^L_2$ to $t_r^L$. Then the probability of this event is 
bounded above by
\be
\sum_{{p^L_1\in\Sigma^*_L:\atop p^L_1\not\in I^L_1}}
\sum_{p^L_2\in\Sigma^*_L}
\exp\{-\W(\cC(p^L_1,p^L_2))\}\leq 
\ee
$$
O(L^2)\min\{\exp\{-\W(\cC(p^L_1,p^L_2))\}\,|\,p^L_1\in
\Sigma^*_L\backslash
I^L_1\,,\,p^L_2\in\Sigma^*_L\} \,.
$$
Suppose that $\cC$ denotes the polygonal line giving the minimum; 
scaled
by $1/L$ we get a polygonal line in $Q$, denoted by $\cC^*$,
from $A$ to some point $P_1^*$, then from $P_1^*$ to $P_2^*$ 
and finally from
$P_2^*$ to $B$. Let $\theta^*$ be the angle between the straight 
line from $A$
to $P_1^*$ with the wall. We have
\be
\W(\cC)=L\W(\cC^*)\geq L(g(\theta^*,a)+g(\theta_Y,b))\,.
\ee
By hypothesis 
\be
|\theta^*-\theta_Y|\geq {1\over L^{1/2}}O((M\log L)^{1/2})\,. 
\ee
Therefore (use a Taylor expansion of $g$ around $\theta_Y$ and 
the monotonicity
of $g(\theta,x)$ on $[0,\theta_Y]$, respectively 
$[\theta_Y,\pi/2]$) there
exists a positive constant $\alpha$ such that
\bea
\W(\cC^*)&\geq & g(\theta_Y,a)+g(\theta_Y,b) +
{\alpha M\log L\over L}\\ 
&=&
\W^*+{\alpha M\log L\over L}\,.\nnm 
\eea
We conclude that the probability, that condition $b_2$ is not 
satisfied,
is bounded above by $O(L^{c+2})/L^{\alpha M}$.
If $M$ is large enough, the second statement of the theorem is
true.  

{\bf 3.} The proof of the third statement of the theorem is similar.
\qed

\subsection{Finite size effects for the correlation length}
\label{correlationlength}

By duality we can interpret Proposition \ref{lowerbd} and
Theorem \ref{mainthm} at temperatures 
above $T_c$. The fact that $t_l^L$ and $t_r^l$ are points on the 
boundary of  $\Lambda_L^*$ is not important for the dual model 
above
the critical temperature, as one can check easily from the proofs
of sections \ref{Estimate} and \ref{Continuum}.

Let $\beta^*<\beta_c$ and ${\Lambda'}_L$ be the
subset of $\bZ^{2*}$ obtained by translating $\Lambda_L^*$
by $(0,-L/2)$, that is,
\be
{\Lambda'}_L:=\{\,t\in\bZ^{2*}:\, t+(0,L/2)\in\Lambda_L^*\,\}\,.
\ee
We consider the Ising model with free b.c.
on ${\Lambda'}_L$ with coupling constants
\be
J^*(t,t'):=\cases{h^*>0& if $t(2)=t'(2)=-1/2-L/2$,\cr
                 1& otherwise.\cr}
\ee
The corresponding Gibbs state is denoted by 
$\bk{\,\cdot\,}_{{\Lambda'}_L}$ or by
$\bk{\,\cdot\,}_{{\Lambda'}_L}(\beta^*,h^*)$. 
When $L\ra\infty$ the states 
$\bk{\,\cdot\,}_{{\Lambda'}_L}(\beta^*,h^*)$
converge to the unique infinite Gibbs state
$\bk{\,\cdot\,}(\beta^*)$ of the model with coupling constant $1$,
independently of the value of $h^*$, since $\beta^*<\beta_c$.
The (horizontal) correlation length $\xi(\beta^*)$ is therefore
independent of $h^*$ and is given by the formula 
\be\label{correl}
\xi(\beta^*):=-\lim_{L\ra\infty}{1\over L}
\log\,\bk{\,\sigma(t)\sigma(t+(L,0))\,}(\beta^*)\,.
\ee
Theorem \ref{mainthm}, as well as Proposition \ref{lowerbd},
show that in general we do not get
the same result if we take the thermodynamical limit and the limit in
(\ref{correl}) together.  Indeed,  we can find 
$h^*$ and $t_L$  such that the distance of $t_L$ to the boundary of
the box ${\Lambda'}_L$ is $O(L)$ and 
\be\label{long}
-\lim_{L\ra\infty}{1\over L}\log\,
\bk{\,\sigma(t_L)\sigma(t_L+(L,0))\,}_{{\Lambda'}_L}(\beta^*,h^*)
\not =\xi(\beta^*)\,,
\ee
because in the random--line representation of the two--point
correlation function the random lines are concentrated near a part
of the boundary of the box ${\Lambda'}_L$. 
Borrowing the terminology
of \cite{SML} about the long--range order, we can say that 
there is no
equivalence in general between the ``short'' correlation length
$\xi(\beta^*)$ and a  ``long'' correlation length like in 
(\ref{long}).  
Proposition \ref{produality} states that this equivalence
holds when $h=h^*=1$, the
correlation length $\xi(\beta^*)$ being equal to the surface 
tension
of an horizontal interface of the dual model. 
The reason for the validity of
Proposition \ref{produality} can be formulated
in physical terms: the dual model is in the complete wetting regime.

\section{Sharp triangle inequality }\label{sti}

The main property of  the surface tension, which we use in this
paper, is the sharp triangle inequality (STI). 
We recall some basic facts about the Wulff shape and prove that
STI is equivalent to the property that the curvature of the Wulff
shape is bounded above.  This slightly extends the result of 
Ioffe \cite{I}. In particular we do not suppose that the
surface tension is differentiable. Our approach is
geometrical. 

\subsection{Convex body and support function}

Let $\stau :\bR^2\ra \bR$, $x\mapsto \stau(x)$, be a positively
homogeneous convex function, which is strictly positive at
$x\not =0$. In this section $\bk{y^*,x}$ denotes the Euclidean
scalar product of $y^*\in\bR^2$ and $x\in\bR^2$.
The conjugate function $\stau^*$,
\be
\stau^*(y^*):=\sup_{x}\{\bk{y^*,x}-\stau(x)\}\,,
\ee
is the indicator function of a convex set $W\subset\bR^2$
defined by $\stau$\footnote{In Statistical Mechanics, when $\stau$ 
is the surface
tension, W is called
the {\bf Wulff shape} and (\ref{wulff}) 
the Wulff construction.},
\be
\stau^*(y^*)=\cases{0& if $y^*\in W$\cr
\infty & otherwise.\cr}
\ee
Because $\stau$ is strictly positive at $x\not =0$, the interior
of $W$ is non--empty ($W$ is a convex body) and contains $0$.
The function $\stau$ is the support function of 
$W$\footnote{$\stau$ is a norm if and only if $W=-W$.},
\be\label{a1}
\stau(x)=\sup_{y^*\in W}\bk{y^*,x}\,.
\ee
Given $x$, we define the half--space $H(x)$,
\be
H(x):=\{y^*:\,\bk{y^*,x}\leq\stau(x)\}\,.
\ee
We have the important relation,
\be\label{wulff}
W=\{x^*\in\bR^2:\,\bk{x^*, y}\leq\stau(y)\;,\; \forall y\not=0\}
=\bigcap_{y\not=0}H(y)\,.
\ee
A pair of points $(y^*,x)\in\bR^2\times\bR^2$ is in 
{\bf duality}\footnote{ Duality of Convexity Theory.}
if and only if 
\bea\label{a2}
\bk{y^*,x}&=&\stau(x)+\stau^*(y^*)\\
&=&\stau(x)\nonumber\,.
\eea
If $(y^*,x)$ are in duality and $x\not=0$, then 
$y^*\in\partial W$, the  boundary of $W$. Moreover, in such a case
$(y^*,\lambda x)$ are in duality for any positive scalar
$\lambda$. In the following $\hat{x}$ is always a unit vector in
$\bR^2$; there is  at least one $x^*\in\partial W$,
which is in duality with $\hat{x}$ for any $\hat{x}$.
The geometrical meaning of $\hat x$ is the following: there is a support plane
for $W$ at $x^*$ normal to $\hat x$.
By convention the pair $(x^*,\hat{x})$ is in duality, so
that  $\bk{x^*,\hat{x}}=\stau(\hat{x})$.
We may have
$y^*_1\not = y_2^*$, such that $(y_1^*,\hat{x})$ and 
$(y_2^*,\hat{x})$ are in  duality. 

\begin{lem}\label{lema1}
1. Suppose that  $y^*_1$ and $y_2^*$ are two different points of
\be
\cF(\hat{x}):=\{y^*\in\bR^2:\,(y^*,\hat{x})\;\hbox{are in
duality}\}\,.
\ee
Then $y^*=\alpha y^*_1 +(1-\alpha)y_2^*\in \cF(\hat{x})$,
for all $\alpha\in[0,1]$.

2. The set $\cF(\hat{x})$
is  equal to the set of subdifferentials of $\stau$ at
$\hat{x}$,
\be
\partial \stau(\hat{x}):=\{y^*\in \bR^2:\,
\stau(z+\hat{x})\geq \stau(\hat{x}) +\bk{y^*,z}\quad
\forall z\in\bR^2\}\,.
\ee
\end{lem}

\prf
1. follows from (\ref{a2}); 2. is proven e.g. in \cite{PV2} 
section 4.1. \qed

The set $\cF(\hat{x})$ of Lemma \ref{lema1} is a {\bf facet} of $W$
with (outward) normal $\hat{x}$ ($0$ belongs to interior of $W$).
Therefore, existence of a facet is
equivalent to non--uniqueness of the subdifferentials of
$\stau$ at $\hat{x}$ or to non differentiability of $\stau$. 
For a given $y^*\in\partial W$ we  may
have two different  vectors $\hat{x}_1$ and $\hat{x}_2$,
such that $(y^*,\hat{x}_1)$ and  $(y^*,\hat{x}_2)$ are in 
duality.
This situation corresponds to the existence of a {\bf corner }
of $W$  at $y^*$.

\begin{lem}\label{lema2}
There is a corner of $W$ at $y^*$ if and only if
there exists a segment
$[\hat{x}_1,\hat{x}_2]:=\{x:\,x=\hat{x}_1+t(\hat{x}_2-\hat{x}_1)
\,,\,t\in[0,1]\}$, with $\hat{x}_1\not =\hat{x}_2$, on which
$\stau$ is affine.
\end{lem}

\prf
Suppose that there is a corner at $y^*$. Then
\bea\label{b1}
\bk{y^*,\hat{x}_1+t(\hat{x}_2-\hat{x}_1)}
&=&
(1-t)\stau(\hat{x}_1)+ t\stau(\hat{x}_2)\\
&\leq&
\sup_{y^*\in W}\bk{y^*,\hat{x}_1+t(\hat{x}_2-\hat{x}_1)}
\nonumber\\
&=&
\stau((1-t)\hat{x}_1+t\hat{x}_2)\,.\nonumber
\eea
Since $\stau$ is convex we have equality in (\ref{b1}).

Suppose that $\stau$ is affine on $[\hat{x}_1,\hat{x}_2]$.
Let $x_{1/2}:=1/2(\hat{x}_1+\hat{x}_2)$. If
$y^*\in\partial \stau(x_{1/2})$, then
$y^*\in\partial \stau(\hat{x}_k)$, $k=1,2$. Indeed, for all $z$
\be
\stau(z)- \stau(x_{1/2})-\bk{y^*,z-x_{1/2}}\geq 0\,.
\ee
But, if $\stau$ is affine on $[\hat{x}_1,\hat{x}_2]$, then
\be
{1\over 2}\sum_{k=1}^2\{\,
\stau(\hat{x}_k)- \stau(x_{1/2})-\bk{y^*,\hat{x}_k-x_{1/2}}\,\}=0
\,.
\ee
Therefore
\be
\stau(\hat{x}_k)= \stau(x_{1/2})+\bk{y^*,\hat{x}_k-x_{1/2}}\,.
\ee
From this it follows that $y^*\in\partial \stau(\hat{x}_k)$,
\be
\stau(z)
\geq  
\stau(x_{1/2})+\bk{y^*,z-x_{1/2}}\\
=
\stau(x_{k})+\bk{y^*,z-\hat{x}_k}\quad\forall z\,,
\ee
which implies in our case that $(y^*,\hat{x}_k)$ are in duality.
\qed

\subsection{Curvature}

We recall the notion of  curvature  of $W$ at $x^*$.
Let $U$ be an open neighbourhood of $x^*$. Let $\cT_i(x^*,U)$ be 
the  family of discs $\cD$ with the following properties
\begin{enumerate}
\item
$\partial\cD$ is tangent\footnote{The precise definition is the following:
there is a common support plane at $x^*$ for $W$ and $\cD$.} to $\partial W$ at 
$x^*$;
\item
$W\cap U\supset \cD\cap U$.
\end{enumerate}
We allow the degenerate cases where the disc is a single
point or a half--plane. Consequently $\cT_i(x^*,U)\not =\emptyset$.
We denote by $\rho(\cD)$ the radius of the disc
$\cD$ and  set
\be
\underline{\rho}(x^*,U):=\sup\{\rho(\cD):\,\cD\in \cT_i(x^*,U)\}\,.
\ee
Clearly $\underline{\rho}(x^*,U_1)\leq \underline{\rho}(x^*,U_2)$
if $U_1\supset U_2$.
The {\bf lower radius of  curvature} at $x^*$ is defined as
\be
\underline{\rho}(x^*):=
\sup\{\underline{\rho}(x^*,U):\, U\;\hbox{open neighb. of
$x^*$}\}\,. 
\ee
Similarly, we introduce
$\cT_s(x^*,U)\not=\emptyset$ as
the  family of discs $\cD$ with the following properties
\begin{enumerate}
\item
$\partial\cD$ is tangent to $\partial W$ at $x^*$;
\item
$W\cap U\subset \cD\cap U$.
\end{enumerate}
We set
\be
\overline{\rho}(x^*,U):=\inf\{\rho(\cD):\,\cD\in \cT_s(x^*,U)\}\,.
\ee
The {\bf upper radius of  curvature} at $x^*$ is defined as
\be
\overline{\rho}(x^*):=
\inf\{\overline{\rho}(x^*,U):\, U\;\hbox{open neighb. of
$x^*$}\}\,. 
\ee
Given $x^*, y^*\in\partial W$, $x^*\not = y^*$,
let $\cC(x^*;\rho_{y^*})$ be the circle 
of radius $\rho_{y^*}$, which is 
tangent to $\partial W$ at $x^*$ and 
goes through $y^*$\footnote{We suppose that $\cC(x^*;\rho_{y^*})$ intersects
the interior of $W$.}.
If $y^*\in U$, then 
\be\label{classic}
\underline{\rho}(x^*,U)\leq \rho_{y^*}\leq
\overline{\rho}(x^*,U)\,.
\ee
If ${\rho}(x^*):=\underline{\rho}(x^*)=\overline{\rho}(x^*)$, then
the {\bf  radius of  curvature} at $x^*$ is ${\rho}(x^*)$ and
the {\bf curvature}  at $x^*$ is $\kappa(x^*):=1/\rho(x^*)$
(see Chapter 1 of \cite{S}).
From (\ref{classic}) we get\footnote{ 
If $\lim_{y^*\ra x^*}\rho_{y^*}=\rho$, then for every $\varepsilon>0$
there exists a neighbourhood $U$ such that for all $y^*\in U$,
$|\rho_{y^*}-\rho|\leq \varepsilon$. Therefore
$\rho-\varepsilon\leq \underline{\rho}(x^*,U)\leq
\overline{\rho}(x^*,U)\leq \rho+\varepsilon$.}
\be
\underline{\rho}(x^*)=\overline{\rho}(x^*)\Longleftrightarrow
\lim_{y^*\ra x^*}\rho_{y^*}=\rho(x^*)\,.
\ee

If $x^*$ is a corner, then $\underline{\rho}(x^*,U)=0$ 
for any open  neighbourhood $U\ni x^*$ 
and $\overline{\rho}(x^*,U)$ is as small as we wish, provided
$U$ is small enough.
Therefore $\rho(x^*)=0$.
However, we may have  $\rho(x^*)=0$ 
when $x^*$ is not a corner, as the following example shows.
We define a convex body by its boundary,
\be
\{\,z^*(1)(t):=\cos(t)|\cos(t)|^{.6}\,,\,
z^*(2)(t):=\sin(t)|\sin(t)|^{.6}\;,\;t\in[0,2\pi]\,\}\,.
\ee
There is a unique support line at every point of the
boundary. At the four points ($t=k\pi/2$, $k=0,\ldots 3$)
it is elementary to verify that $\rho(x^*)=0$. 

\begin{lem}\label{curvature}
Let  $W$ be a convex compact body such that its lower radius of curvature
is bounded below uniformly by $K_0>0$.
Then, given  $\rho< K_0$, there is  a circle $\cC(x^*;\rho)\subset W$ 
of radius $\rho$,  which is tangent to $ \partial W$ at $x^*$ for
any $x^*$.
\end{lem}

\prf
Since the lower radius of curvature is positive, there is no corner.
Consequently, for any $y^*\in\partial W$
there is unique $\hat{y}$ in duality with $y^*$. The hypothesis also
implies that at every $x^*\in\partial W$ there is a disk $\cD(x^*)$
with the properties: the radius $\rho(\cD(x^*))$ of $\cD(x^*)$ is non-zero, 
$\cD(x^*)\subset W$ and $\partial \cD(x^*)$ is tangent to 
$\partial W$ at $x^*$. Since $W$ is convex, the convex envelope
of all these discs is a subset of $W$. Therefore, since $W$ is compact,
we can find $\delta>0$, such that $\rho(\cD(x^*))\geq \delta$ for any $x^*$.

Let $x^*\in\partial W$ and $\hat{y}$ be given. Let 
$\cD(x^*,\hat{y})\subset H(\hat{x})\cap H(\hat{y})$
be the largest disc, which is tangent to $\partial H(\hat{x})$
at $x^*$. If $\hat{x}=\hat{y}$, then the radius $r(x^*,\hat{y})$
of $\cD(x^*,\hat{y})$ is infinite, otherwise it is finite. 
Since $\stau(\,\cdot\,)$ is continuous, $r(x^*,\hat{y})$ is a 
continuous function of $\hat{y}$ at any $\hat{y}\not =\hat{x}$.
We set
\be
r(x^*):=\inf_{\hat{y}}r(x^*,\hat{y})\,.
\ee
Let $\{\hat{y}_n\}$ be a minimizing sequence, such that
$\lim_n r(x^*,\hat{y}_n)=r(x^*)$ and $\lim_n \hat{y}_n=:\hat{y}$.
There are two cases: $\hat{y}=\hat{x}$ and $\hat{y}\not=\hat{x}$.

If $\hat{y}=\hat{x}$, then  $r(x^*)\geq K_0$.
Suppose the converse, $r(x^*)<K_0$.
Then, for any $n$ such that $r(x^*,\hat{y}_n)<K_0$, we can find a
disc $\cD_n$ and a neighbourhood $U_n$ of $x^*$, such that $\cD_n$
is tangent to $\partial W$ at $x^*$ and
\be
W\cap U_n\supset \cD_n\cap U_n\supset  \cD(x^*,\hat{y}_n)\cap U_n\,.
\ee
Let  $z^*_n$ be the point of contact of $\cD(x^*,\hat{y}_n)$ with 
$\partial H(\hat{y}_n)$.  Since 
$ W$ is convex,  $\partial W$
intersects $\partial \cD(x^*,\hat{y}_n)$ at some point $t^*_n$
belonging to the circle arc of  $\partial \cD(x^*,\hat{y}_n)$
from $x^*$ to $z^*_n$. Since $r(x^*,\hat{y}_n)<K_0$ and 
$\hat{x}=\lim_n\hat{y}_n$, we also have $\lim_n z^*_n=x^*$
and thus $\lim_n t^*_n=x^*$. But this contradicts (\ref{classic}).
Thus $r(x^*)\geq K_0$;  for any $\rho< K_0$ 
there is  a circle $\cC(x^*;\rho)$ 
of radius $\rho$,  which is tangent to $ \partial W$ at 
 $x^*\in\partial W$;   $\cC(x^*;\rho)\subset W$ by (\ref{wulff}).

If $\hat{y}\not =\hat{x}$, then $r(x^*,\hat{y})=r(x^*)\geq \delta$ and the
disc $\cD(x^*,\hat{y})\subset W$  by (\ref{wulff}).
Let $r:=\inf_{x^*}r(x^*)$;
we claim that $r\geq K_0$. Suppose the converse, $r< K_0$.
Let $\{z^*_n\}$ be a minimizing sequence such that $r(z^*_n)<K_0$,
$\lim_n r(z^*_n)=r$ and $\lim_n z^*_n=:z^*$. 
For every $n$ there exists $\hat{y}_n$
such that $r(z^*_n,\hat{y}_n)=r(z^*_n)$ and
$\cD(z^*_n,\hat{y}_n)$ is the largest disc in $W$, which is tangent
to $\partial W$ at $z^*_n$. Since $r< K_0$, there exists $\hat{y}\not=\hat{z}$,
so that 
\be
r=r(z^*,\hat{y})\,.
\ee
Indeed, if $\cD(z^*)\subset W$ is a disc tangent to $\partial W$ at 
$z^*$,
then by convexity the convex envelope of $\cD(z^*)$ and 
$\cD(z^*_n,\hat{y}_n)$ is a subset  of $W$. If $\rho(\cD(z^*))>r$, then
the discs $\cD(z^*_n,\hat{y}_n)$ are
not the largest discs in $W$ which are tangent to $\partial W$ at $z^*_n$,
when $n$ is sufficiently large. The existence of $\cD(z^*,\hat{y})$
and the convexity of $W$ imply the existence of an open set $U$, such that
\be
\partial W\cap U=\partial\cD(z^*,\hat{y})\cap U\,.
\ee
Indeed, $\cD(z^*,\hat{y})$ is tangent to $\partial W$ at $z^*$ and also 
at some $y^*$ in duality with $\hat{y}$; moreover, at any point $x^*\in\partial
W$ there exists a disc of radius $r$ contained in $W$, tangent to $\partial W$ at $x^*$.
But this contradicts $r< K_0$.
\qed

\subsection{STI}

\begin{pro}\label{stiequivalence}
Let $W$ be a convex compact body 
and $\stau$ be its support function. Then the
following statements are equivalent.

\begin{enumerate}
\item
The lower radius of curvature of $\partial W$ is uniformly bounded
below by $K_0>0$.
\item
There exists 
a positive constant $K_1$ such that for any
$\hat{x}$ and $\hat{y}$ 
\be\label{a3}
\bk{x^*-y^*,\hat{x}}\geq K_1 \|\hat{x}-\hat{y}\|^2\,.
\ee
\item
There exists a positive constant $K_2$ such that for any
$x$, $y\in\bR^2$
\be\label{a4}
\stau(x)+\stau(y)-\stau(x+y)\geq
K_2(\|x\|+\|y\|-\|x+y\|)\,. 
\ee
\end{enumerate}
\end{pro}

{\bf Remarks:} 1. Suppose that the curvature is bounded above everywhere
by $\kappa$. Then 1. holds with $K_0=1/\kappa$. 
1. implies 2. with $K_1=1/2\kappa$;  this follows by modifying
slightly the proof given below: if $\bk{x^*-y^*,\hat{x}}\leq 4K_1$,
then there exists $\hat{v}$ such that 
$\bk{x^*-y^*,\hat{x}}=2K_1\|\hat{x}-\hat{v}\|^2$ and 
$\bk{\hat{x},\hat{y}}\geq \bk{\hat{x},\hat{v}}$.
2. implies 3. with $K_2=1/\kappa$. 

2. The validity of the sharp triangle inequality
implies absence of corner for $W$,
since it prevents $\stau$ to be affine on segments
$[\hat{x}_1,\hat{x}_2]$ with $\hat{x}_1\not =\hat{x}_2$.
However, the  example before Lemma \ref{curvature}
shows that the converse is not true.

\prf
We prove  $1\Rightarrow 2$. Let $x^*,y^*\in
\partial W$, $x^*\not = y^*$ and $0< 2K_1<K_0$. The circle $\cC(x^*;2K_1)$
of radius $2K_1$, center $d^*$, which is tangent to
$\partial W$ at $x^*$, is a subset of $W$ (Lemma \ref{curvature}). 
If 
\be
\bk{x^*-y^*,\hat{x}}\geq 2 K_1\,,
\ee
then for any $\hat{y}$ 
\be
\bk{x^*-y^*,\hat{x}}\geq K_1/2\|\hat{x}-\hat{y}\|^2\,.
\ee
Suppose that
\be
\bk{x^*-y^*,\hat{x}}\leq 2K_1\,.
\ee
We can find $z^*\in \cC(x^*;2K_1)$ such that, if
$\hat{v}:=(z^*-d^*)/\|z^*-d^*\|$ and $\varphi$ is the angle between
the unit vectors $\hat{x}$ and $\hat{v}$, then
$\bk{\hat{x},\hat{v}}\geq 0$ and
\be
\bk{x^*-y^*,\hat{x}}=2K_1(1-\cos \varphi)=K_1\|\hat{x}-\hat{v}\|^2\,.
\ee
We claim that 
\be
\bk{\hat{x},\hat{y}}\geq \bk{\hat{x},\hat{v}}\,.
\ee
Suppose the converse.  Let $\cC(y^*;2K_1)\subset W$ be the circle 
of radius $2K_1$, which is tangent to $\partial W$ at $y^*$.
By hypothesis  the line perpendicular to $\hat{v}$ at $z^*$ and
the support line  at $y^*$ perpendicular to $\hat{y}$ intersects
at an interior point of $H(\hat{x})$; this implies that 
$\cC(y^*;2K_1)\not \subset H(\hat{x})$, which is in contradiction
with $W\subset H(\hat{x})$.
Therefore
\be
\|\hat{x}-\hat{y}\|\leq \|\hat{x}-\hat{v}\|\,
\ee
and 
\be
\bk{x^*-y^*,\hat{x}}= K_1\|\hat{x}-\hat{v}\|^2
\geq K_1\|\hat{x}-\hat{y}\|^2\,.
\ee

We prove  $2\Rightarrow 1$. Suppose that 
\be
\bk{x^*-y^*,\hat{x}}
\geq K_1\|\hat{x}-\hat{y}\|^2\,.
\ee
Let $\cC(x^*;\rho_{y^*})$ be the circle of  radius
$\rho_{y^*}$,  which is  tangent to $\partial W$ at $x^*$ and 
goes through $y^*$; let $c^*$ be its center  and 
$\hat{u}:=(y^*-c^*)/\|y^*-c^*\|$. Assume furthermore that
$\bk{\hat{x},\hat{u}}\geq 0$. Let $v:=\hat{x}+\hat{u}$ and $\hat{v}=v/\|v\|$.
Then
\be\label{aaa}
{\rho_{y^*}\over 2}\|\hat{x}-\hat{u}\|^2
=
\bk{x^*-y^*,\hat{x}}\,
\ee
and $\bk{x^*-y^*,\hat{v}}=0$.
Since $\partial W$ is convex, there exists $z^*$
``between'' $x^*$ and $y^*$ such that $\hat{z}=\hat{v}$  and
\be\label{edf3}
\|\hat{x}-\hat{z}\|\leq\|\hat{x}-\hat{y}\|\,.
\ee
On the other hand,
\be\label{edf1}
\|\hat{x}-\hat{u}\|\leq 
\|\hat{x}-\hat{v}\|+\|\hat{v}-\hat{u}\|\,
\ee
and
\be\label{edf2}
\|\hat{x}-\hat{v}\|=\|\hat{v}-\hat{u}\|=\|\hat{x}-\hat{z}\| \,.
\ee
If $\hat{x}=\hat{y}$, then $\rho_{y^*}=\infty$; otherwise,
using (\ref{aaa}) to (\ref{edf2}),  
\bea
2\rho_{y^*}\|\hat{x}-\hat{v}\|^2
&\geq &
{\rho_{y^*}\over 2}\|\hat{x}-\hat{u}\|^2\\
&= & 
\bk{x^*-y^*,\hat{x}}\nonumber\\
&\geq&
K_1\|\hat{x}-\hat{y}\|^2\nonumber\\
&\geq&
K_1\|\hat{x}-\hat{v}\|^2\,.\nonumber
\eea
Since this holds for any $y^*$ in a neighbourhood of $x^*$,
we have $\underline{\rho}(x^*)\geq K_1/2$.

We prove  $2\Rightarrow 3$. We set
\be
z:=x+y\quad,\quad z^*:=(x+y)^*\quad,\quad
\hat{z}:={x+y\over\|x+y\|}\,. 
\ee
We  have, using (\ref{a1}) and (\ref{a2}),
\bea\label{a5}
\stau(x)+\stau(y)-\stau(z)&=&\bk{x^*,x}+\bk{y^*,y}
-\bk{z^*,z}\\
&=&\bk{x^*-z^*,x}+\bk{y^*-z^*,y}\nonumber\\
&=&\|x\|\,\bk{x^*-z^*,\hat{x}}+\|y\|\,\bk{y^*-z^*,\hat{y}}
\nonumber\,.
\eea 
By elementary trigonometry  
\be\label{a6}
\|x\|+\|y\|-\|z\|={1\over 2}\left(
\|x\|\cdot\|\hat{x}-\hat{z}\|^2+
\|y\|\cdot\|\hat{y}-\hat{z}\|^2\right)\,, 
\ee
so that (\ref{a3}) implies (\ref{a4}).

We prove  $3\Rightarrow 2$.
Let $(x^*,\hat{x})$ and
$(y^*,\hat{y})$ be given; we set $z:=\hat{x}+\hat{y}$.
Using (\ref{a6}), $\|\hat{x}-\hat{z}\|=\|\hat{y}-\hat{z}\|$
and $\|\hat{x}-\hat{y}\|\leq \|\hat{x}-\hat{z}\|+
\|\hat{z}-\hat{y}\|$ we have
\bea
\bk{x^*-y^*,\hat{x}}&=&\bk{x^*,\hat{x}}+\bk{y^*,\hat{y}}
-\bk{y^*,\hat{x}+\hat{y}}\\
&\geq &
\stau(\hat{x})+\stau(\hat{y})-\stau(z)\nonumber \\
&\geq &
K_2(\|\hat{x}\|  +\|\hat{y}\|-\|z\|)\nonumber\\
&=&K_2\|\hat{x}-\hat{z}\|^2\nonumber\\
&\geq&{K_2\over 4}\|\hat{x}-\hat{y}\|^2\nonumber\,.
\eea
\qed

\pagebreak


\end{document}